\RequirePackage{rotating}
\documentclass[fleqn, usenatbib]{mnras}

\usepackage[T1]{fontenc}

\DeclareRobustCommand{\VAN}[3]{#2}
\let\VANthebibliography\thebibliography
\def\thebibliography{\DeclareRobustCommand{\VAN}[3]{##3}\VANthebibliography}

\usepackage{graphicx}	
\usepackage{amsmath}	
\usepackage{amssymb}	

\usepackage{flushend}
\usepackage{balance}

\usepackage{float}
\usepackage{caption}
\usepackage{lipsum}

\usepackage{fixltx2e}
\usepackage{xcolor}
\usepackage{tablefootnote}

\usepackage{pdflscape}
\usepackage{longtable}
\usepackage{booktabs}
\usepackage{color}
\usepackage{amssymb}
\usepackage{mathtools}
\usepackage{xspace}
\usepackage{rotating}
\usepackage{appendix}
\usepackage{multirow}
\usepackage{multicol}
\usepackage{array}
\usepackage{subcaption}
\usepackage{comment}
\usepackage{tabularx}
\usepackage{bigstrut}
\usepackage{threeparttable}
\usepackage{afterpage}
\usepackage{capt-of}

\usepackage[normalem]{ulem}
\usepackage{siunitx}

\newcommand{\ecyc}[1]{\ensuremath{E_{\rm{C}}}}

\newcommand{\krtext}{\textcolor{black}}
\newcommand{\krrtext}{\textcolor{black}}

\usepackage{newtxtext, newtxmath}

\title[Eclipse flares in HMXBs]{Flares during Eclipses of High Mass X-ray Binary Systems Vela X-1, 4U 1700-37, and LMC X-4}

\author[Rikame et al.]{
Ketan Rikame,$^{1,2}$\thanks{E-mail: rikame.bhaskar@res.christuniversity.in}
Biswajit Paul,$^{2}$
Rahul Sharma,$^{2}$
V. Jithesh,$^{1}$ and
KT Paul$^{1}$\thanks{Deceased}
\\
$^{1}$CHRIST (Deemed to be University), Department of Physics and Electronics, Bangalore 560029. Karnataka, India\\
$^{2}$Raman Research Institute, Astronomy and Astrophysics, C.V. Raman Avenue, Bangalore 560080. Karnataka, India\\
}

\date{Accepted XXX. Received YYY; in original form ZZZ}

\pubyear{2021}

\begin{document}
\label{firstpage}
\pagerange{\pageref{firstpage}--\pageref{lastpage}}
\maketitle

\begin{abstract}

In eclipsing X-ray binary systems, the direct X-ray emission is blocked by the companion star during the eclipse. We observe only reprocessed emission that contains clues about the environment of the compact object and its chemical composition, ionization levels, etc. \krtext{We have found flares in some X-ray binaries during their eclipses.} The study of eclipse flares provides additional clues regarding the size of the reprocessing region and helps distinguish between different components of the X-ray spectrum observed during the eclipse. In the archival data, we searched for flares during eclipses of high-mass X-ray binaries and found flares in three sources: Vela X-1, LMC X-4, and 4U 1700-37. Comparing spectral properties of the eclipse flare and non-flare data, we found changes in the power-law photon index in all three sources and multiple emission lines in Vela X-1 and 4U 1700-37. The fluxes of prominent emission lines showed a similar increase as the overall X-ray flux during the eclipse flare, suggesting the lines originate in the binary environment and not in the interstellar medium. We also observed a soft excess in 4U 1700-37 that remains unchanged during both eclipse flare and non-flare states. Our analysis suggests that this emission originates from the extremely thin shell of the stellar wind surrounding the photosphere of its companion star. \krtext{The detection of short (100-200 seconds) count-rate doubling timescale in 4U 1700-37 and LMC X-4 indicates that the eclipse reprocessing occurs in a region larger than, but comparable to the size of the companion star.}

\end{abstract}



\section{Introduction}

X-ray binary systems comprise of a companion star and a compact object in orbit, where the matter transfer takes place from the companion star to the compact object. The compact object can be a neutron star or a black hole. X-rays originate close to the compact object due to matter accretion from the companion star. These X-rays are called primary X-rays. Some part of these X-rays come out of the system directly. But a significant fraction interacts with the surrounding matter producing secondary X-rays. This phenomenon is known as X-ray reprocessing and the secondary X-rays are called reprocessed X-rays. In eclipsing X-ray binary systems, during eclipse, the primary X-rays are blocked by the companion star and only reprocessed emission is observed making them ideal systems to study the reprocessed emission. In this work we focus on eclipsing high-mass X-ray binary (HMXB) systems. X-ray flares, that are often observed in HMXBs can also be found during eclipse. \citep{islam2016,martinezchicharro2021}. The X-ray flares (hereinafter referred to as just "flares") may originate near the compact star due to increase in the rate of matter accretion onto the compact object. This can be driven by the inhomogeneities in the stellar wind from the companion star or any instability in the accretion disk. The X-ray source is blocked from the direct view by the companion star during the eclipse. Thus the flares observed during the eclipse must be due to scattering of light into the observer's line of sight by the matter around the binary. Study of the flares during the eclipse give additional clues regarding size of the reprocessing region, and also help distinguish between different components of the X-ray spectrum observed during the eclipses. \krtext{We searched for variability, indicative of potential flares, during eclipses of High-Mass X-ray Binaries (HMXBs) in a large volume of archival X-ray data. Within this data, observations from three sources, namely Vela X-1 (1 observation), LMC X-4 (1 observation), and 4U 1700-37 (2 observations) showed flares during eclipse.} In this work, we discuss spectral properties of the eclipse flare and eclipse non-flare \footnote[1]{The terms "eclipse non-flare" and "eclipse persistent emission" are used in a similar context in this paper.} data, in these sources.


\begin{figure*}
  \centering
    \includegraphics[keepaspectratio=true, height=9.0cm, angle =0]{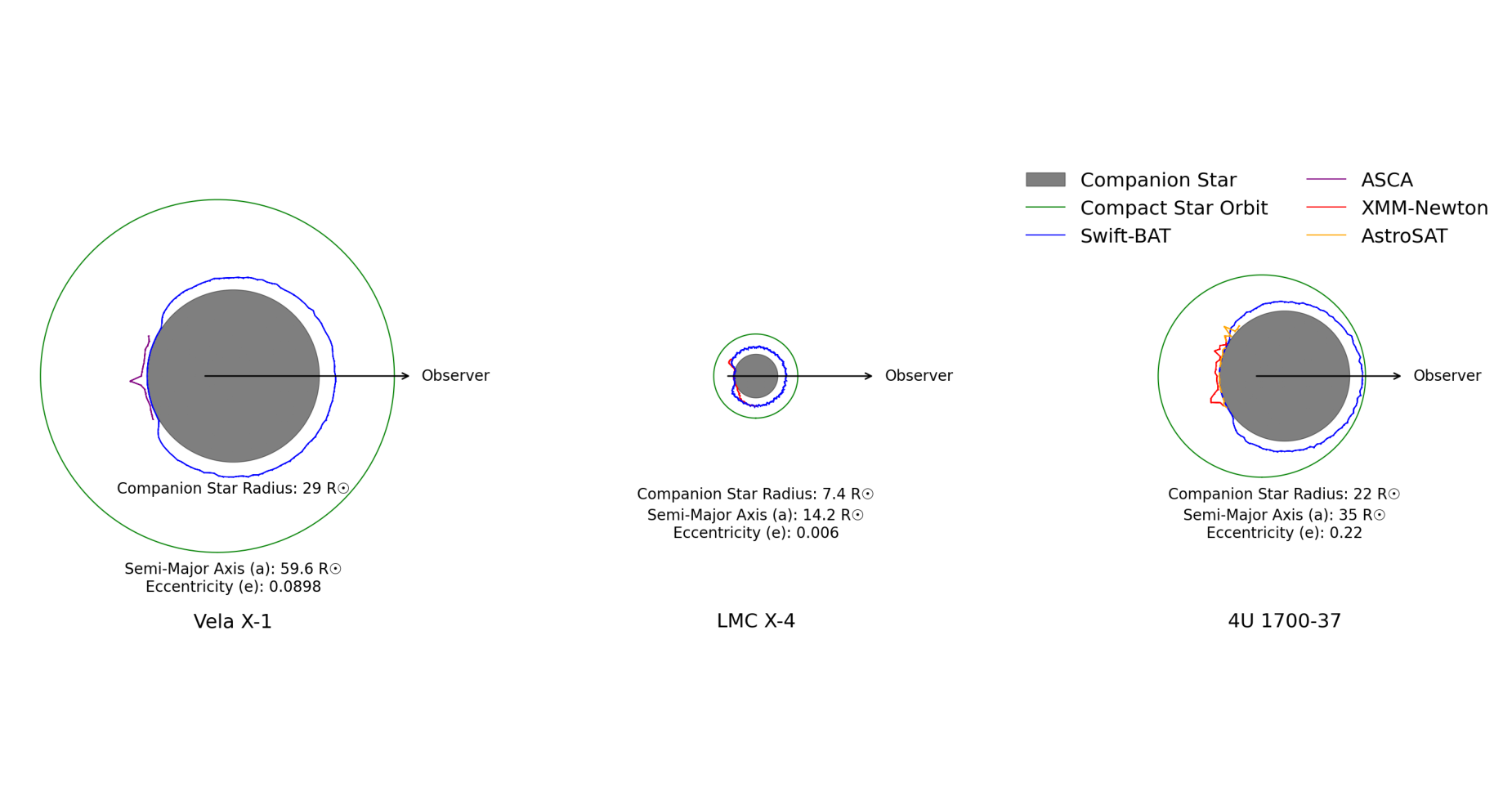}
    \caption{A to-scale sketch of Vela X-1, LMC X-4, and 4U 1700-37 is presented. The companion star is represented as a grey circle, while the orbit of the compact star around the companion is shown in green color. The \textit{swift}-BAT orbital profile, displayed in blue, is plotted radially over the companion star's surface. It is plotted in a linear scale with arbitrary normalization, serving as an indicator of the system's overall brightness at various radial locations of the compact star in its orbit around the companion star. The light curves of the observations analyzed in this work are folded with their respective orbital periods and plotted similarly to the \textit{swift}-BAT orbital profile (Please refer to the legend for details).
} \label{binary_to_scale_vis}
\end{figure*}

\krtext{A to-scale sketch of these three systems, illustrating the orbital arrangement of the compact star around its companion star, is displayed in Figure \ref{binary_to_scale_vis}. The light curves of the observations analyzed in this work, along with respective \textit{swift}-BAT orbital profile, plotted radially over the companion star's surface are also shown in the figure as an indicator of the system's overall brightness at various radial locations of the compact star in its orbit around the companion star.}

Vela X-1 (4U 0900-040) is an eclipsing HMXB system discovered by \cite{chodil1967} with a rocket observation. The companion star in Vela X-1, HD 77581, is a Supergiant star of mass $26\pm1$ $M_{\odot}$ \citep{falanga2015}, and the compact object is a neutron star of mass $1.77_{-0.08}^{+0.08}$ $M_{\odot}$ \citep{rawls2011}. According to \cite{falanga2015}, the estimated radius of HD 77581 is $29\pm1$ $R_{\odot}$, and the semi-major axis of the binary orbit is $59.7\pm7$ $R_{\odot}$. The neutron star accretes matter from the stellar wind of the companion star. The spin period of the neutron star is about 283 seconds \citep{mcclintock1976} and the orbital period of the system is about 8.9 days \citep{ulmer1972}, with the neutron star being eclipsed for about two days. The source is located at a distance of 2 kpc \citep{nagase1989} and shows average intrinsic X-ray luminosity of $\sim$ 10\textsuperscript{36}erg s\textsuperscript{-1} \citep{watanabe2006}. Intense X-ray variability observed in Vela X-1 is attributed to the strongly structured wind of the optical companion. The source exhibits both, giant flares in which the flux increases by more than the factor of 10, and off states in which the flux drop to less than the 10\% of it's average flux \citep{kreykenbohm1999,kreykenbohm2008}. The X-ray continuum emission spectrum has been modelled either by a absorbed power law with a high energy cut-off \citep{nagase1986,kreykenbohm1999,maitra2013} or by a negative positive exponential (NPEX) model \citep{kreykenbohm2002}. Soft excess observed below 3 keV can be explained by emission from diffuse gas excited by collisions and photoionization \citep{hickox2004} and it has been modelled by \cite{haberl1994} as thermal bremsstrahlung with \textit{kT} $\sim$ 0.5 keV. The remarkably line dominated eclipse spectrum observed with \textit{ASCA} shows lines of He-like ions of neon, magnesium and silicon, in addition to an intense 6.4 keV iron emission line \citep{nagase1994}. Cyclotron resonance scattering features (CRSFs) have been observed at $\sim$ 25 keV and $\sim$ 50 keV by \cite{kreykenbohm2002} using the data from the Rossi X-ray Timing Explorer (RXTE). \cite{orlandini1998} found the CRSF only around 55 keV using \textit{BeppoSAX} data and debated the detection at 25 keV \citep{orlandini2006}. In recent years, the CRSFs have been clearly detected at both $\sim$ 25 keV and $\sim$ 50 keV, using data from multiple observatories such as \textit{Suzaku} \citep{maitra2013}, \textit{INTEGRAL} \citep{wangw2014}, \textit{NuSTAR} \citep{diez2022}, and \textit{HXMT} \citep{liug2022}. The typical pulse profile of Vela X-1 shows complex five peak structure at low energies and a double sinusoidal pattern at high energies \citep{nagase1989}. The shape of the pulse profile remains similar during normal emission state and flares. However, during the low state, the pulsation amplitude reduces significantly \citep{choi1996}. \cite{choi1996} also observed modulation in the iron line energy with the pulse phase during the flare.

LMC X-4 is a wind and disk fed persistent accretion powered eclipsing HMXB in the Large Magellanic Cloud first detected with the Uhuru satellite \citep{giacconi1972}. LMC X-4 harbours a $1.25_{-0.10}^{+0.11}$ $M_{\odot}$ neutron star and a $14.5_{-1.0}^{+1.1}$ $M_{\odot}$ O8 giant companion \citep{meer2007}. \cite{falanga2015} estimated the radius of the companion star to be $7.4\pm4$ $R_{\odot}$ and the semi-major axis of the binary orbit to be $14.2\pm2$ $R_{\odot}$. It is an eclipsing binary pulsar with a pulse period of $\sim$ 13.5 seconds \citep{white1978,kelly1983} and orbital period of $\sim$ 1.4 day \citep{kelly1983,ilovaisky1984} with an eclipse duration of $\sim$ 5 hours \citep{white1978}. Along with the spin period and the orbital motion, a super-orbital variability of 30.4 d is observed \citep{lang1981,paulkitamoto2002,molkov2015}. Orbital period decay with a timescale of $\sim$ 10\textsuperscript{6} yr has been reported by \cite{naikpaul2004}. For an assumed distance of 50 kpc \citep{pietrzy2013}, LMC X-4 shows unabsorbed continuum X-ray luminosity of $\sim$ 1$\times$10\textsuperscript{38}erg s\textsuperscript{-1} during the high state of the super-orbital period \citep{neilsen2009}, and during the low state, the flux can be lower by a factor of upto 60 \citep{naikpaul2004}. The source exhibits frequent flares during which the luminosity can reach up-to 10\textsuperscript{39}erg s\textsuperscript{-1} \citep{kelly1983,levine1991,beri2017,brumback2020}. The X-ray spectrum has been historically modelled by hard power law with a high-energy cutoff and a soft excess, modified with a line of sight absorption and a strong iron emission line. The soft excess in LMC X-4 is explained by reprocessing of the hard X-rays by optically thick accreting material \citep{bpaul2002,hickox2004} and it has been modelled by thermal Comptonization \citep{labarbera2001}, blackbody radiation \citep{naikpaul2002}, and thermal bremsstrahlung \citep{naikpaul2004}. The pulse profile is observed to change in both shape and phase during flares \citep{levine2000,beri2017,shtykovsky2018}. The variation of soft spectral component and the power law with the pulse phase shows phase difference indicating different origin of soft and hard X-rays \citep{beri2017}. Quasi periodic oscillations (QPOs) with frequencies ranging from 1 to 2 mHz have been observed in the flares of LMC X-4 by \cite{moon2001}. Additionally, a QPO with a frequency of $\sim$ 26 mHz was observed during persistent emission state of LMC X-4 \citep{rikame2022,sharma2023}. From the QPOs detected in the persistent emission data of LMC X-4, \cite{rikame2022} estimated the magnetic field of the neutron star in LMC X-4 to be $\sim$ 30$\times$10\textsuperscript{12} Gauss.

4U 1700-37 is a wind fed HMXB discovered with the Uhuru observatory \citep{jones1973}. The orbital period of about 3.4 days with an eclipse duration of about 1.1 days has been observed by \cite{jones1973}. \cite{islam2016} have estimated the orbital period decay rate ($ \dot{P}/P $) of the system to be around $\sim$ 5$\times$10\textsuperscript{7} yr\textsuperscript{-1}. The nature of the compact object is not yet fully understood because of the absence of pulsations that can be attributed to the weak magnetic field or the alignment of magnetic axis with the rotation axis of the compact object \citep{reynolds1999}. 4U 1700-37 is powered by a dense stellar wind from the supergiant HD 153919, the hottest and most luminous companion known in HMXBs \citep{jones1973}. The radius of HD 153919 is $21.9_{-0.5}^{+1.3}$ $R_{\odot}$ \citep{clark2002}. \cite{hainich2020} estimated the mass loss rate ($\dot{M}$) of HD 153919 to be 2.51$\times$10\textsuperscript{-6} $M_{\odot}$ yr\textsuperscript{-1} and terminal velocity of the stellar wind ($\nu$\textsubscript{$\infty$}) to be 1.9$\times$10\textsuperscript{3} km s\textsuperscript{-1}. 4U 1700-37 is located at a distance of $\sim$1.7 kpc \citep{bailerjones2018} and displays a luminosity of $\sim$ 3$\times$10\textsuperscript{35}erg s\textsuperscript{-1} during the quiescence state \citep{martinezchicharro2018}. The source has been observed to show intense flares and substantial variability, which is thought to be caused due to accretion from the supergiant's inhomogeneous wind \citep{white1983}. The X-ray continuum spectrum can be modelled with an absorbed power law and high energy cut-off \citep{haberl1989,reynolds1999}. Soft excess observed below 4 keV has been previously modelled with a thermal bremsstrahlung \citep{haberl1992} as well as with a blackbody \citep{meer2005}. In the eclipse, eclipse egress, and low flux \textit{XMM-Newton} spectra, \cite{meer2005} has reported many recombination and fluorescent lines, including highly ionised and fluorescent Fe K\textsubscript{$\alpha$} emission lines. The lines are most prominent during eclipse. \textit{Chandra} - High Energy Transmission Gratings (HETGs) observation of 4U 1700-37 during an X-ray eclipse showed flares and X-ray spectrum showed multiple emission lines \citep{martinezchicharro2021}. They found emission-line brightness from K-shell transitions to be correlated with continuum illumination. However, the emission lines are still significant during eclipse which can be explained if K$\alpha$ emission comes from the bulk of the wind. Highly ionized lines diminish during the eclipse indicating their origin to be close to the compact object.

In this paper, we report detection of X-ray flares during the eclipses in several sources and discuss the spectral properties of the eclipse flare and eclipse non-flare states in the above-mentioned sources. \krtext{The paper is structured as follows. Section \ref{ana_res} provides a description of the archival data survey (Section \ref{observations_sec}), eclipse variability search  (Section \ref{eclipse_fl_id}), the data reduction procedures (Section \ref{data_red}), and the X-ray instruments used in this analysis (Section \ref{instruments}). The analysis and results are detailed in Section \ref{analysis}. We discuss the implications of X-ray flare studies during eclipses in Section \ref{discussion}, while Section \ref{conclusion} summarizes and concludes the work.}

\section{\krtext{Archival Data Survey, OBSERVATIONS, DATA REDUCTION and X-ray instruments}}
\label{ana_res}

\subsection{\krtext{Archival Data Survey}}
\label{observations_sec}

We shortlisted six eclipsing HMXBs, viz.: Cen X-3, LMC X-4, SMC X-1, Vela X-1, 4U 1538-522 and 4U 1700-37. We looked for count-rate variability in all available eclipse observations of these six eclipsing HMXBs in the archival X-ray data from Advanced Satellite for Cosmology and Astrophysics \textit{(ASCA)}, Beppo Satellite per Astronomia a raggi X (\textit{BeppoSAX}), Chandra X-ray observatory (\textit{Chandra}), Nuclear Spectroscopic Telescope Array \textit{(NuSTAR)}, Rossi X-ray Timing Explorer \textit{(RXTE)}, \textit{Suzaku}, X-ray Multi-Mirror Mission \textit{(XMM-Newton)} and \textit{AstroSat}.

\subsection{\krtext{Search for variability during eclipse}}
\label{eclipse_fl_id}

\begin{figure*}
  \centering
  \begin{subfigure}[b]{0.4\textwidth}
    \includegraphics[keepaspectratio=true, height=6.0cm, angle =0]{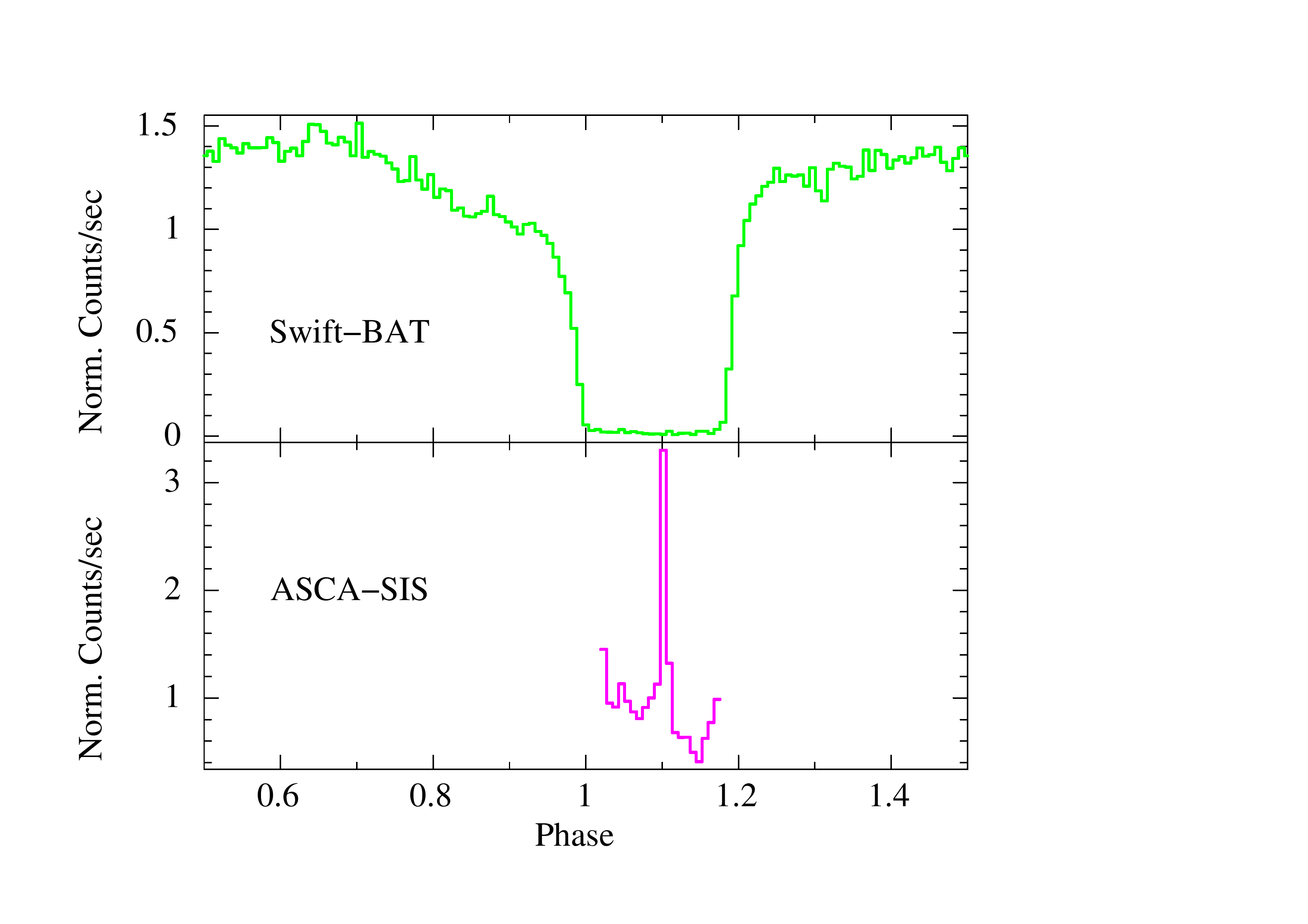}
    \caption{Vela X-1 (ASCA OBSID:43032000)} \label{vela_x_1_orb_prof}
  \end{subfigure}
  \quad
  \begin{subfigure}[b]{0.4\textwidth}
    \includegraphics[keepaspectratio=true, height=6.0cm, angle =0]{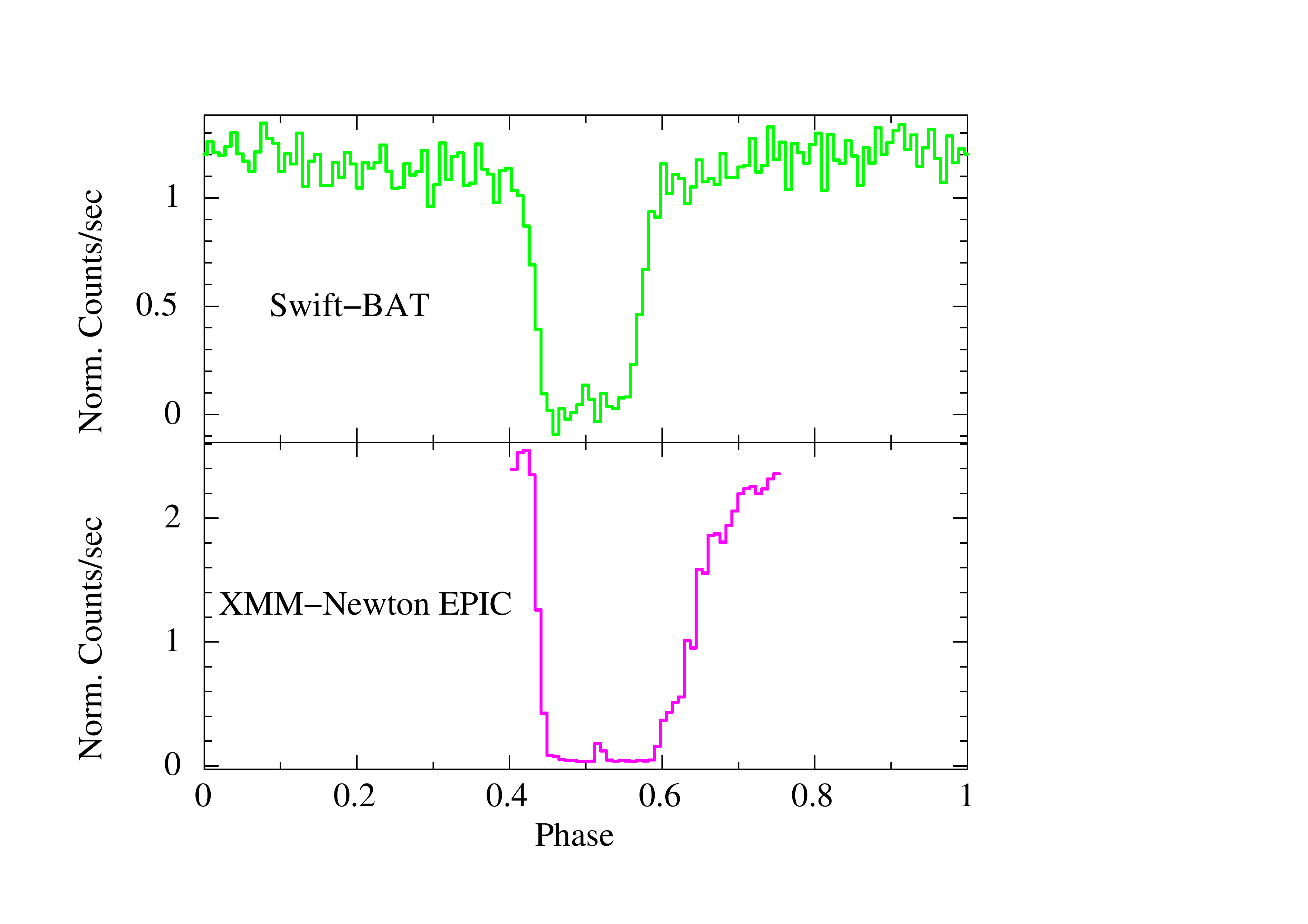}
    \caption{LMC X-4 (XMM-Newton OBSID:203500201)} \label{lmc_x_4_orb_prof}
  \end{subfigure}
  \quad
  \begin{subfigure}[b]{0.4\textwidth}
    \includegraphics[keepaspectratio=true, height=6.0cm, angle =0]{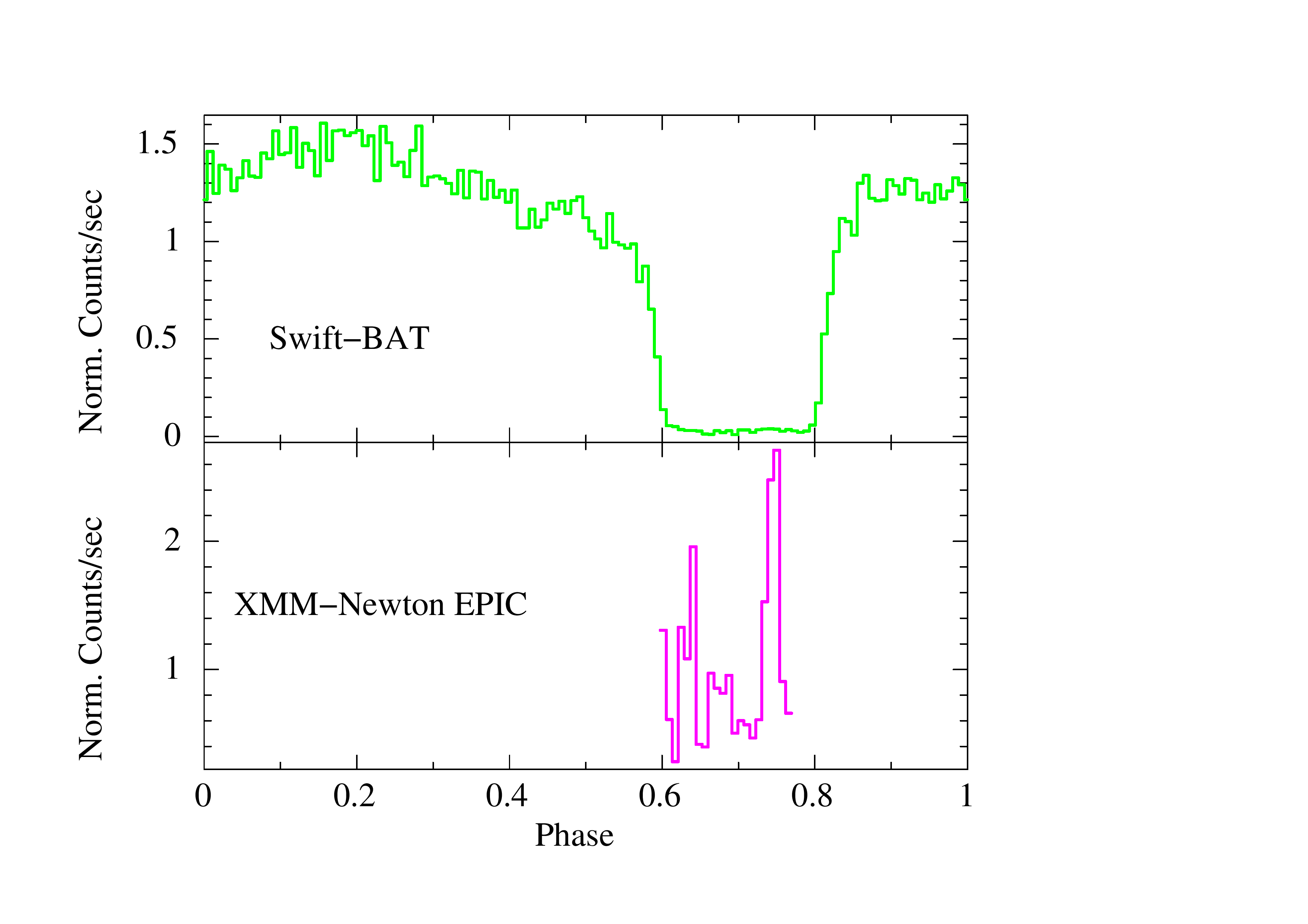}
    \caption{4U 1700-37 (XMM-Newton OBSID:600950101)} \label{4u_1700_37_xmm_orb_prof}
  \end{subfigure}
  \quad
  \begin{subfigure}[b]{0.4\textwidth}
    \includegraphics[keepaspectratio=true, height=6.0cm, angle =0]{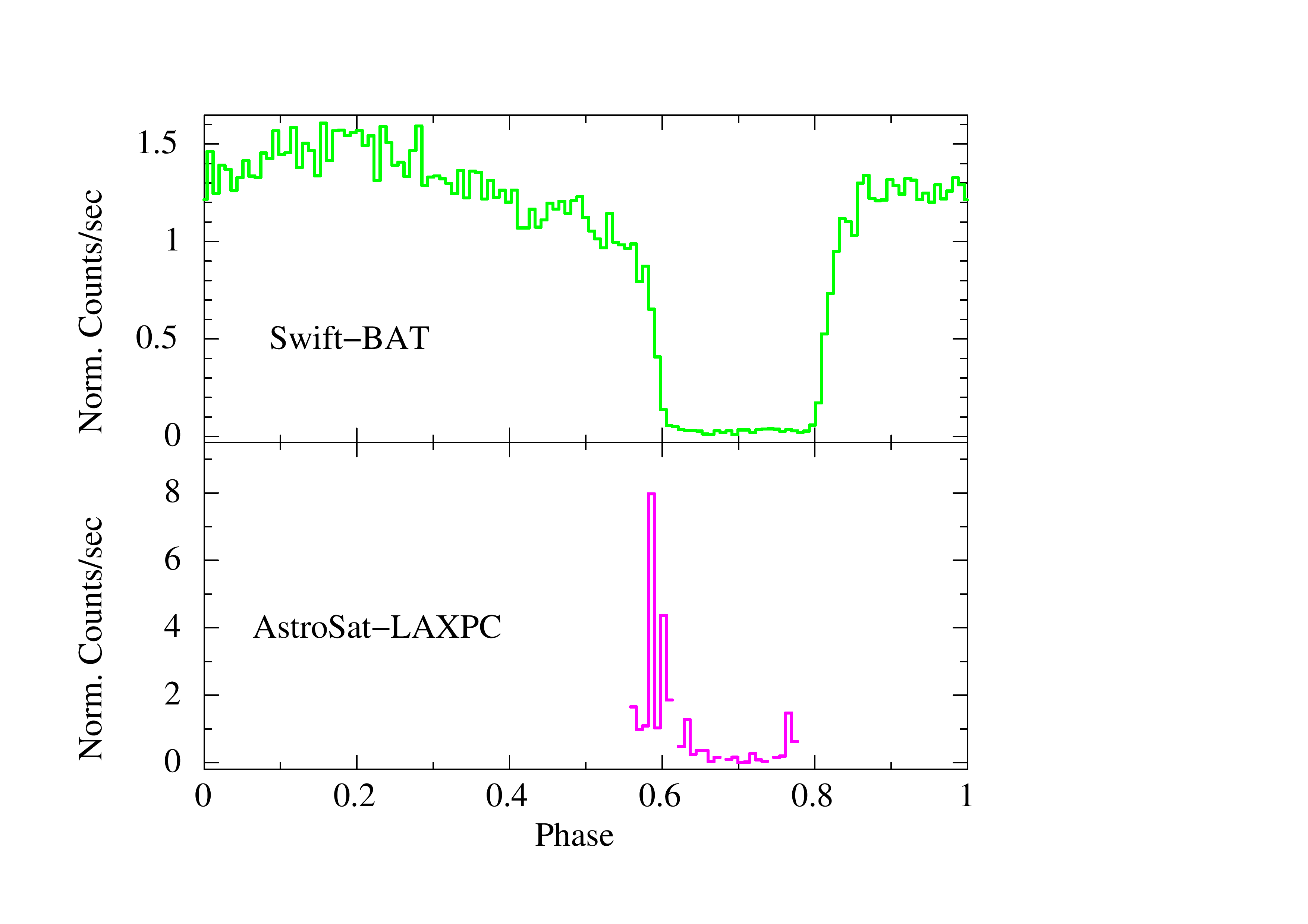}
    \caption{4U 1700-37 (AstroSat OBSID:9000001892)} \label{4u_1700_37_laxpc_orb_prof}
  \end{subfigure}
  \caption{The light curves of the observations showing flares during eclipse, folded with respective orbital periods. In each sub-figure, the top panel shows the 15-50 keV long term average \textit{swift}-BAT orbital profile of the source and the bottom panel shows the light curve of the observation folded with its orbital period. The corresponding source name and observation ID are mentioned in the caption below each sub-figure.}
  \label{orb_prof_plot}
\end{figure*}

\krtext{The eclipse observations were identified by comparing the readily available light curve of each observation with the orbital profile of the source. In cases where readily available data products were not present in the archive, standard data products were extracted using basic data extraction procedures, similar to those described in Section \ref{data_red}.} The orbital profile was obtained by folding the 15-50 keV long term \textit{swift}-BAT source light curve at the orbital period with the help of FTOOL-efold \citep{ftools1999}. The orbital period was determined using FTOOL-efsearch \citep{ftools1999}. Figure \ref{orb_prof_plot} shows the light curves of the observations showing variability during eclipse, folded with respective orbital periods. 

We looked for variability during eclipse, in all the observations, which were either completely or at least partly in eclipse. An eclipse variability was found in \textit{ASCA} observation of Vela X-1 (OBSID: 43032000). LMC X-4 showed a variability during eclipse in one of the \textit{XMM-Newton} observation (OBSID: 203500201). In 4U 1700-37, large variability was observed in \textit{XMM-Newton} (OBSID: 600950101) and \textit{AstroSat} (OBSID: 9000001892) eclipse duration data. In the case of 4U 1700-37, the variability during the eclipse was also observed in the data from \textit{Chandra} (OBSID 18951). However, the observation has already been previously analyzed by \cite{martinezchicharro2021}. Thus we have only analyzed the \textit{XMM-Newton} and \textit{AstroSat} data of 4U 1700-37 here. Table \ref{tab:eclipse_flare_obsid} lists the observations of eclipsing HMXBs showing variability during eclipse, that were analyzed for this work.

\begin{table*}
\begin{center}
\caption{List of observations of eclipsing HMXBs showing flares during eclipse}
\label{tab:eclipse_flare_obsid}
\begin{tabular}{c c c c c}
\hline
Source	&	Observatory 	&	OBSID   &   Mode   &   Date of observation\\ 
	&	&	&  &   yyyy-mm-dd\\ 
\hline
\hline
Vela X-1	&	\textit{ASCA}	&	43032000    &   Bright   &   1995-11-30   \\
LMC X-4	&	\textit{XMM-Newton}	&	203500201    &   Timing   &   2004-06-16   \\
4U 1700-37	&	\textit{XMM-Newton}	&	600950101   &    Imaging   &   2009-09-01   \\
4U 1700-37	&	\textit{AstroSat}	&	9000001892    &   EA   &   2018-02-13   \\
\hline
\end{tabular}
\label{obsid}
\end{center}
\end{table*}

\subsection{\krtext{Observations and data reduction}}
\label{data_red}

Vela X-1 observation by \textit{ASCA} on 30\textsuperscript{th} November 1995 (OBSID 43032000) showed a variability during the eclipse. The Bright mode data from both SIS detectors was combined and used for the analysis. The identification of the eclipse flares and the persistent emission region was conducted as described in Section \ref{light_curves}, and the respective data were extracted and analyzed using FTOOL-\textit{Xselect} \citep{ftools1999}. Source data was extracted from a circular region of radius 2 arcmin centered around the source. Background data was extracted from an annular region around the source of inner(outer) radius 2(4) arcmin. \textit{XMM-Newton} observation of LMC X-4 on 16\textsuperscript{th} June 2004 (OBSID 203500201) showed a variability during the eclipse. In the case of 4U 1700-37 as well, the variability during the eclipse was observed in the data from \textit{XMM-Newton}  acquired on 1\textsuperscript{st} September 2009 (OBSID 600950101). Standard Science Analysis System (SAS 18.0.0) was used to process \textit{XMM-Newton} observation data files (ODF). Procedures given in the online analysis thread were followed \footnote[2]{\url{https://www.cosmos.esa.int/web/xmm-newton/sas-threads}}. Less than half of the incoming flux reaches MOS detectors since part of the flux goes to RGS spectrometers. Moreover, overall count rate in the eclipse data is low due to blocking of the primary radiation by the companion star. Thus, the data from only PN detector is analyzed. LMC X-4 (OBSID 203500201) was observed in the timing mode which enables faster data readout. In the timing mode data of LMC X-4, source data was extracted from a box region with 27$\leq$RAWX$\leq$47 and the background data was extracted from a box region with 3$\leq$RAWX$\leq$5. 4U 1700-37 (OBSID 600950101) was observed with \textit{XMM-Newton} in imaging mode. In the imaging mode data of 4U 1700-37, source data was extracted from a circular region of radius 1 arcmin centered around the source. Background data was extracted from an annular region around the source of inner (outer) radius 1 (2) arcmin. \textit{AstroSat} observation of 4U 1700-37 made on 13\textsuperscript{th} February 2018 also showed variability during eclipse (OBSID $G08\_041T01\_9000001892$). We have used the data from only one of the proportional counter detectors (LAXPC20) due to gas leakage, variable gain and high background observed in other two detectors (LAXPC10 and LAXPC30) \citep{agrawal2017, antia2017, antia2021}. The Event Analysis (EA) mode data from LAXPC20 was processed by using the LAXPC software (LaxpcSoft: version 3.4.2)\footnote[3]{\url{https://www.tifr.res.in/~astrosat_laxpc/LaxpcSoft.html}}. The light curves and spectra for the source and background were extracted from level 1 files by using the tool \textit{laxpcl1}. In all four observations, flare and persistent emission data during the eclipse were extracted using the appropriate \textit{good time interval (GTI)} file.\\
\krtext{The possibility of obtaining simultaneous data with better spectral resolution, from other instruments on-board the observatories where the observations were conducted, was explored to improve spectral analysis. In the case of the \textit{AstroSat} observation of 4U 1700-37, the simultaneous SXT data does not cover the durations of the eclipse flares. For Vela X-1, the GIS detectors on-board \textit{ASCA} have lower energy resolution compared to SIS detectors. For both 4U 1700-37 and LMC X-4, simultaneous spectroscopy data from the RGS instrument on-board \textit{XMM-Newton} is available. Using SAS 18.0.0 and following the procedures outlined in the online analysis guide \textsuperscript{\textcolor{blue}{2}}, RGS spectra were extracted. However, the eclipse duration, already being photon-poor, resulted in poor statistical quality in the spectra and, thus, were not further analyzed.}

\subsection{\krtext{X-ray instruments}}
\label{instruments}

\textit{ASCA} was launched in February 1993, by Japan Aerospace Exploration Agency (JAXA). There are four identical grazing incidence X-ray telescopes (XRT) on-board \textit{ASCA}. Two Solid State Imaging Spectrometers (SIS0 and SIS1) are situated at the focal plane of two of the XRTs and two Gas Imaging Spectrometers (GIS2 and GIS3) are situated at the focal plane other two XRTs. SIS has a superior energy resolution (E/$\Delta$E of $\thicksim$50 at 6 keV) compared to GIS (E/$\Delta$E of $\thicksim$13 at 6 keV). However, GIS offers higher detection efficiency above $\thicksim$3 keV than SIS. SIS has a square field of view (fov) of 20 arcmin x 20 arcmin, whereas GIS offers larger circular fov with diameter of 50 arcmin. \citep{tanaka1994}

\textit{XMM-Newton} launched by European Space Agency (ESA) in December 1999, has two types of X-ray instruments on-board, European Photon Imaging Camera (EPIC) and Reflection Grating Spectrometer (RGS). The EPIC consists of one PN detector \citep{struder_2001} and two MOS detectors \citep{turner_2001}. RGS consists of two spectrometers \citep{herder_2001}. The observatory has three co-aligned telescopes. The PN detector is at the focal plane of one of the telescopes and receives the complete incoming flux. In other two telescopes, via reflection gratings, part of the flux (44\% of the original incoming flux) reaches the MOS detectors and part of the flux is diverted to RGS \citep{lumb_2012}. The MOS and PN cameras have fov of 30 arcmin, in the energy range of 0.2–12 keV, with a spectral resolution of 20–50 (E/$\Delta$E), and angular resolution of 6 arcsec. EPIC can be operated in various modes, namely full frame, large window, small window, and Timing mode, depending upon the observation requirement. Time resolution of 0.03 ms can be obtained with PN camera when operated in timing mode. To provide simultaneous optical and UV coverage, \textit{XMM-Newton} also has co-aligned Optical/UV Monitor (OM) telescope on-board along with above-mentioned X-ray instruments \citep{mason2001}.

\textit{AstroSat} is an Indian multi-wavelength astronomy satellite launched by Indian Space and Research Organisation in September 2015 \citep{agrawal2006,singh2014}. \textit{AstroSat} has five scientific instruments on-board namely, – Scanning Sky Monitor \citep[SSM: 2.5–10 keV; ][]{ramadevi2017,ramadevi2018}, Soft X-ray Telescope \citep[SXT: 0.3–8.0 keV; ][]{singh2014}, Large Area X-ray Proportional Counters \citep[LAXPC: 3–80 keV; ][]{agrawal2017}, Cadmium Zinc Telluride Imager \citep[CZTI: 20–100 keV; ][]{rao2017}, and Ultra-Violet Imaging Telescope \citep[UVIT: 1300–5500 {\AA}; ][]{tandon2017}. \textit{AstroSat}-LAXPC offers a time resolution of 10 $\mu$s and a broad spectral coverage between 3-80 keV. It has a collimator with a 1\textdegree $\times$ 1\textdegree field of view. Three co-aligned units of proportional counter detectors have a very large total effective area (6000 cm\textsuperscript{-2} at 15 keV). The detector with the most stable gain among the three detectors i.e. LAXPC20 has spectral resolution of about 20\% at 30 keV \citep{antia2021}.

The Burst Alert Telescope \citep[BAT; ][]{barthelmy2005} on-board Neil Gehrels Swift Observatory \citep[\textit{swift}; ][]{gehrels2004} is a coded mask aperture instrument with a wide fov of 100\textdegree $\times$ 60\textdegree and Cadmium Zinc Telluride (CZT) detector operating in the energy range of 15-150 keV. We have used long term archival 15-50 keV BAT light curves \citep{krimm2013} to obtain the orbital profiles and identify eclipse phases of the eclipsing HMXBs.




\section{\krtext{ANALYSIS}}
\label{analysis}

\subsection{\krtext{Timing Analysis}}
\label{light_curves}

\begin{figure*}
  \centering
  \begin{subfigure}[b]{0.4\textwidth}
    \includegraphics[keepaspectratio=true, height=6.0cm, angle =0]{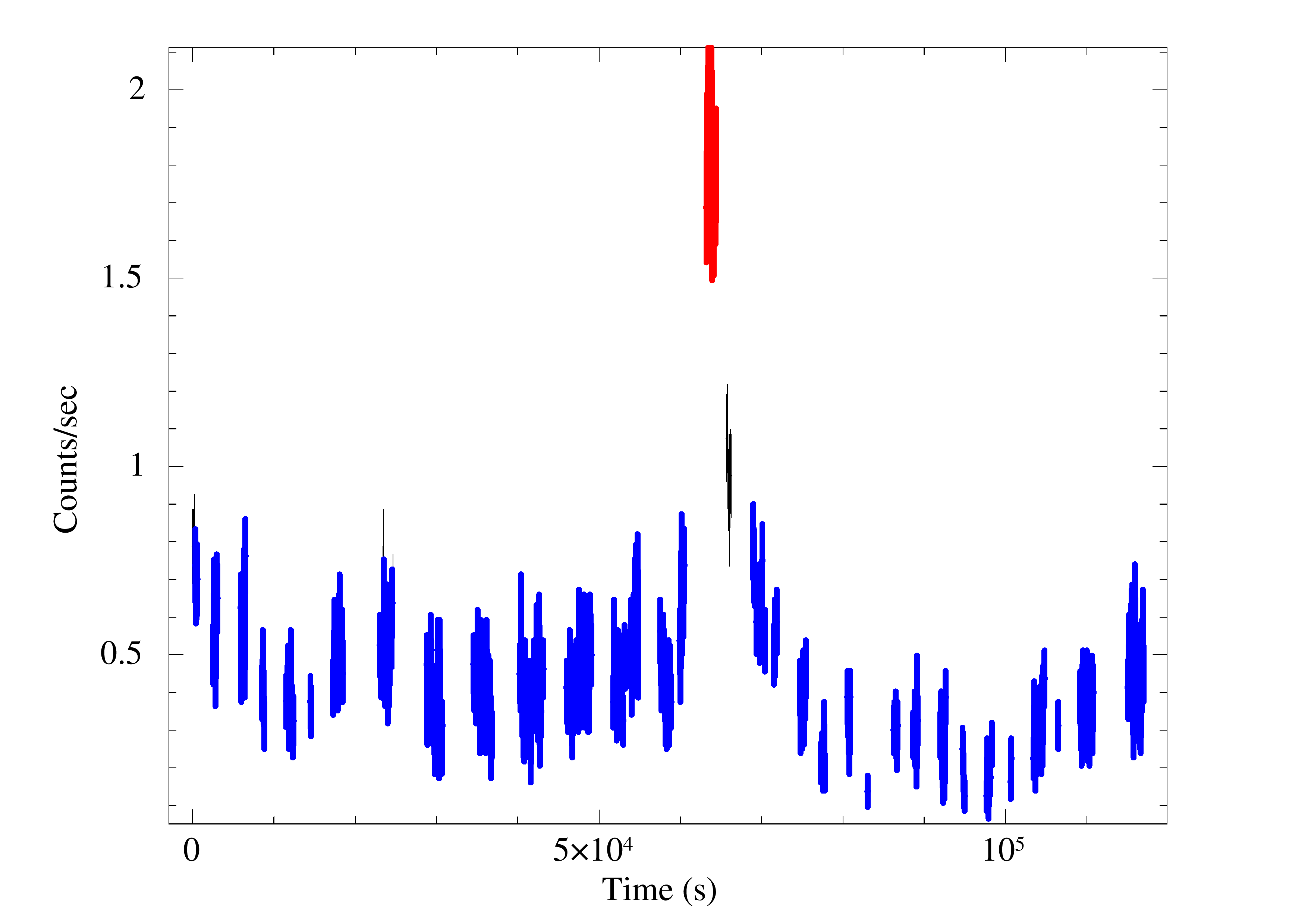}
    \caption{Vela X-1 (ASCA OBSID:43032000)} \label{vela_x_1_lc}
  \end{subfigure}
  \quad
  \begin{subfigure}[b]{0.4\textwidth}
    \includegraphics[keepaspectratio=true, height=6.0cm, angle =0]{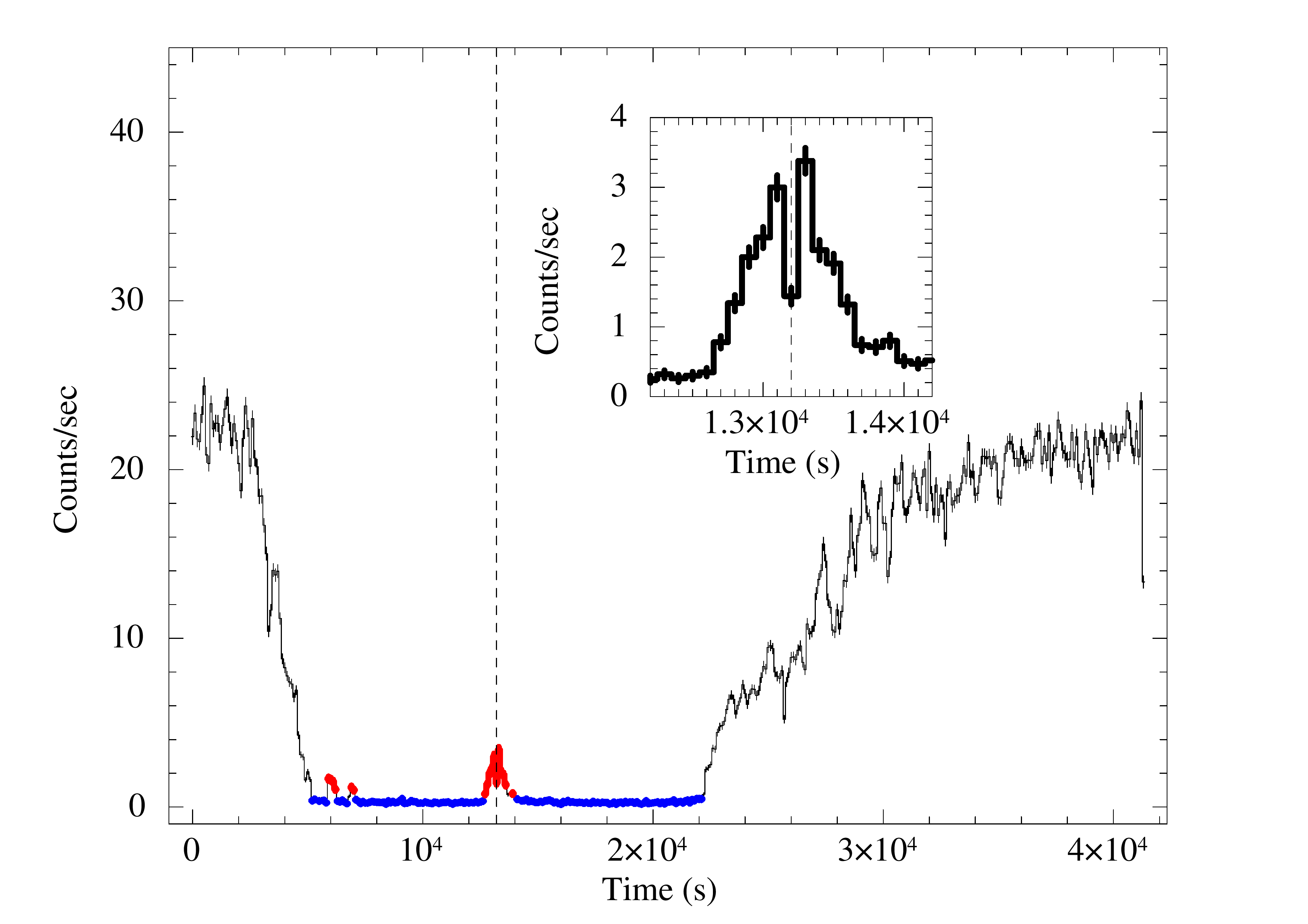}
    \caption{LMC X-4 (XMM-Newton OBSID:203500201)} \label{lmc_x_4_lc}
  \end{subfigure}
  \quad
  \begin{subfigure}[b]{0.4\textwidth}
    \includegraphics[keepaspectratio=true, height=6.0cm, angle =0]{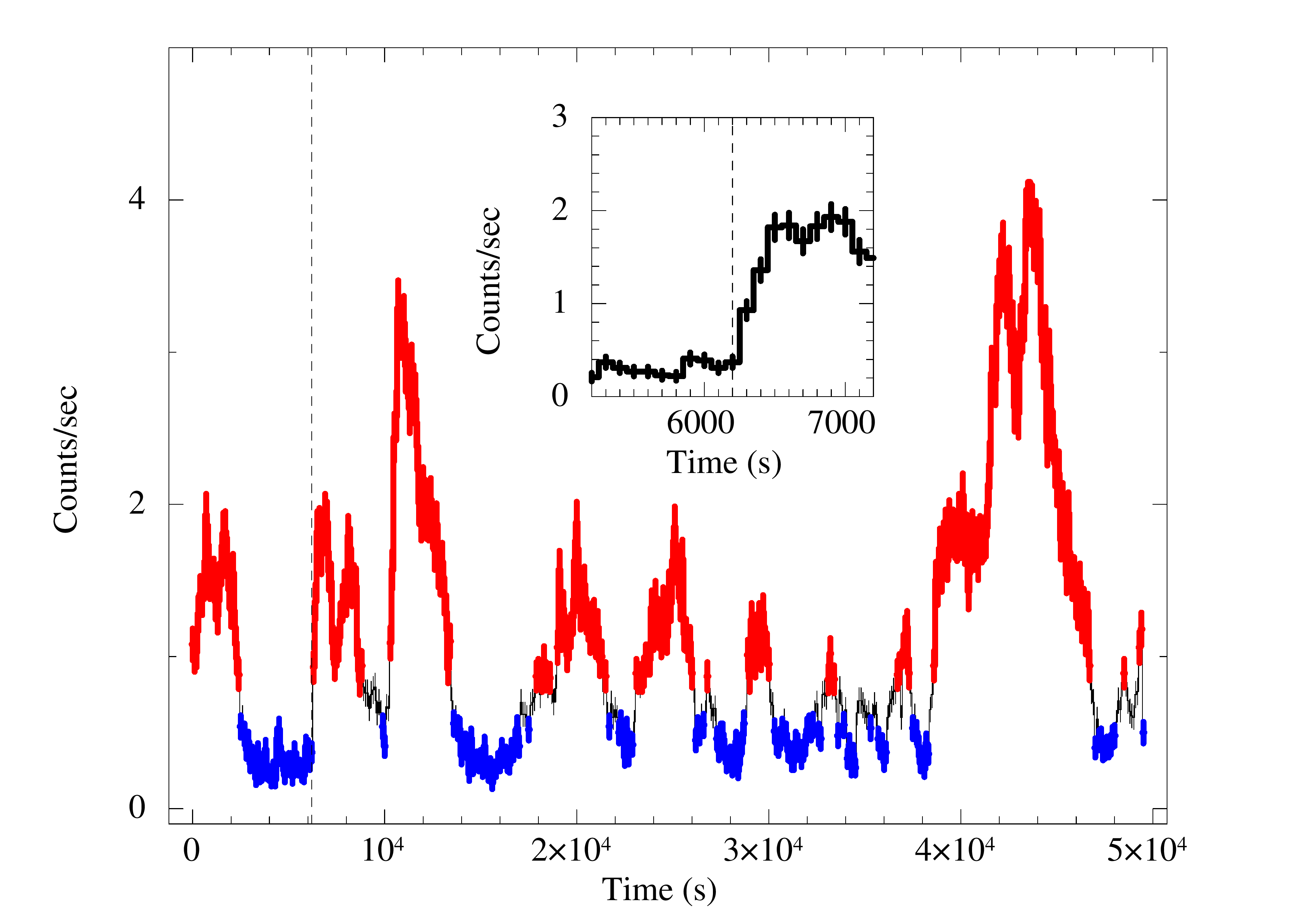}
    \caption{4U 1700-37 (XMM-Newton OBSID:600950101)} \label{4u_1700_37_xmm_lc}
  \end{subfigure}
  \quad
  \begin{subfigure}[b]{0.4\textwidth}
    \includegraphics[keepaspectratio=true, height=6.0cm, angle =0]{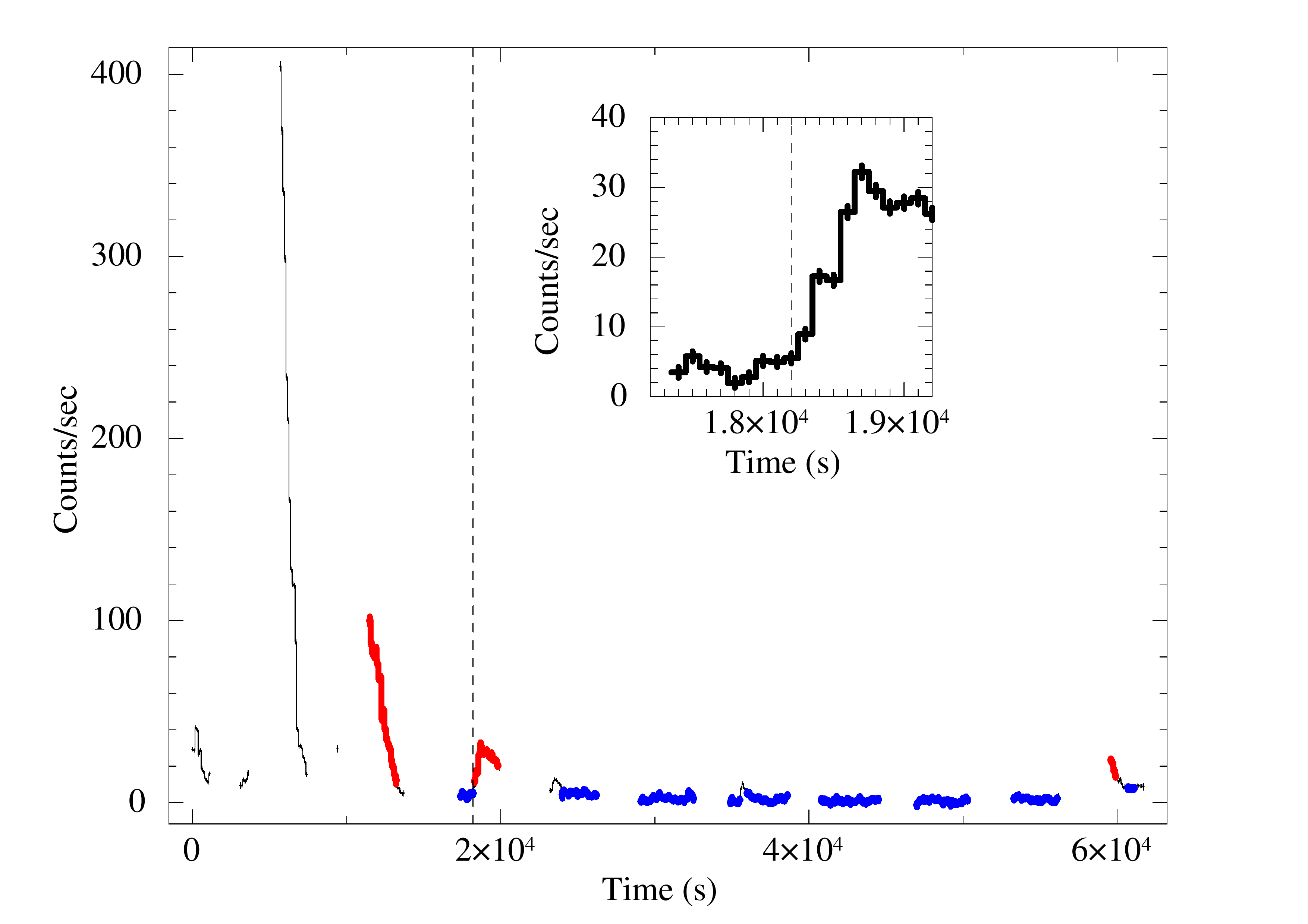}
    \caption{4U 1700-37 (AstroSat OBSID:9000001892)} \label{4u_1700_37_laxpc_lc}
  \end{subfigure}
  \caption{The source light curves \krtext{(Binsize: 100 seconds)} of the observations showing flares during eclipse. The duration highlighted in red shows the eclipse flare data and the duration highlighted in blue shows the eclipse persistent emission data. The unhighlighted duration (shown in black) is not considered for the analysis due to either data being out-of-eclipse or data showing intermediate variability make it difficult to identify between flare and persistent emission duration. \krtext{The inset plots display data points near the eclipse flare. The black dashed line highlights the point with the minimum count rate doubling time (either 100 seconds or 200 seconds) in both the inset plot and the main light curve plot. The count rate doubling time in the Vela X-1 eclipse flare exceeds 1000 seconds, thus it is not highlighted.} The corresponding source name and observation ID are mentioned in the caption below each sub-figure.}
  \label{lc_plot}
\end{figure*}

Figure \ref{lc_plot} shows the source light curves of the observations showing flares during eclipse. \krtext{The eclipse persistent emission count rate was estimated by fitting a constant to the source light curve's eclipse persistent emission duration. To mitigate the effects of statistical variability, any duration up to twice this count rate was identified as eclipse persistent emission, while any duration with a count rate greater than three times this value was considered an eclipse flare. In the light curve plots, the duration highlighted in red corresponded to the eclipse flare data, and the duration highlighted in blue corresponded to the eclipse persistent emission data. The unhighlighted duration (shown in black) was not considered for the analysis due to the data being either out-of-eclipse, during eclipse ingress or egress, or showing an intermediate count rate, making it difficult to distinguish between flare and persistent emission durations.}

\krtext{To assess the variability timescales in the eclipse flare light curves, we calculated the time required to double the current count rate for each data point in 100 seconds binned light curves. The minimum count rate doubling time of 100 seconds was observed near the eclipse flare in \textit{XMM-Newton} observations of LMC X-4 (OBSID:203500201) and 4U 1700-37 (OBSID:600950101). In \textit{AstroSat} observation of 4U 1700-37 (OBSID:9000001892), the minimum count rate doubling time is 200 seconds near the eclipse flare. In the case of the eclipse flare in \textit{ASCA} observation of Vela X-1 (OBSID:43032000), the count rate doubling time exceeds 1000 seconds. The inset plots in Figure \ref{lc_plot} highlight the data points near the eclipse flare with the lowest count rate doubling time.}

\subsection{Spectral analysis}
\label{spec_ana}

\begin{figure*}
  \centering
  \begin{subfigure}[b]{0.4\textwidth}
    \includegraphics[keepaspectratio=true, height=6.0cm, angle =0]{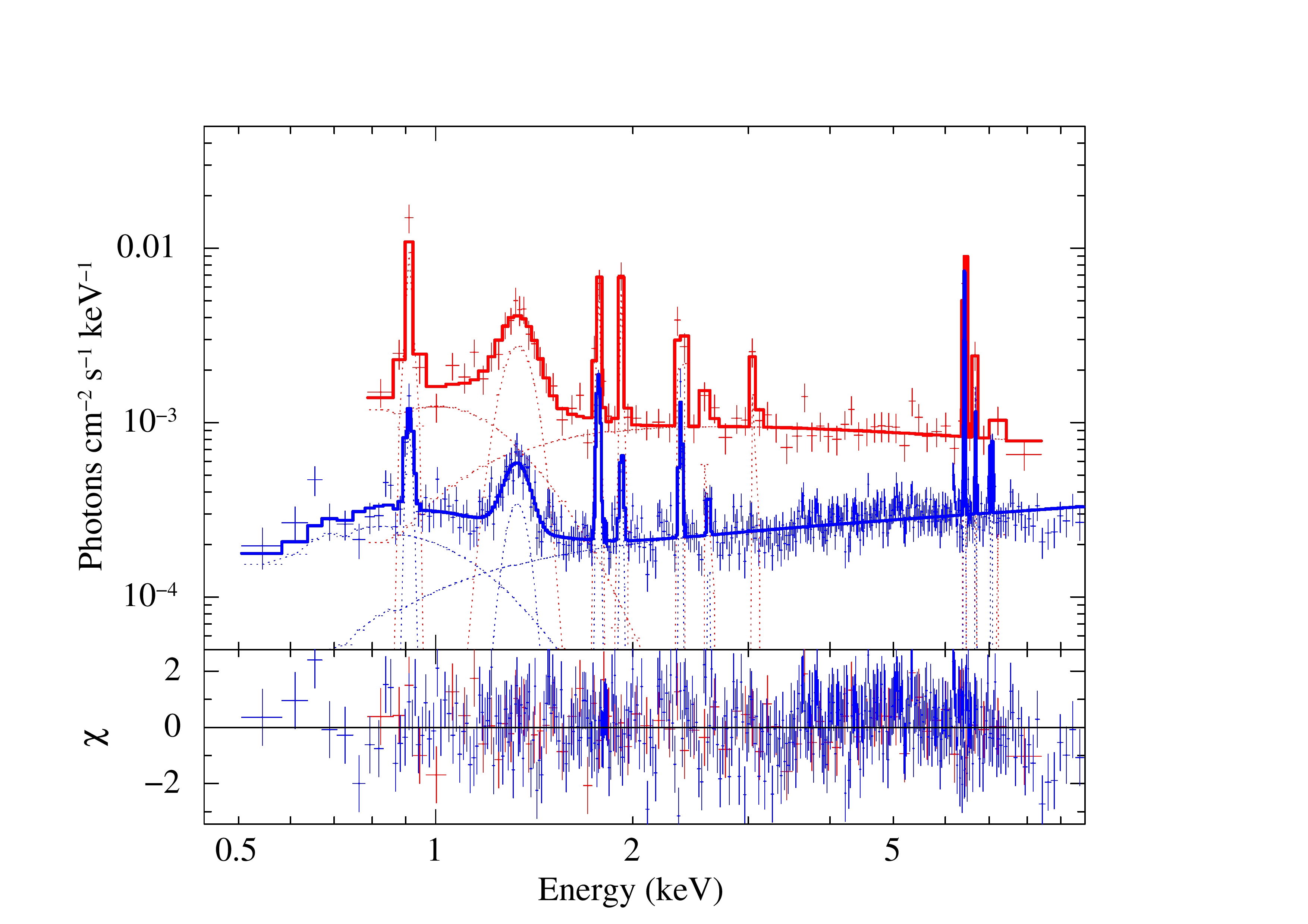}
    \caption{Vela X-1 (ASCA OBSID:43032000)} \label{vela_x_1_spec}
  \end{subfigure}
  \quad
  \begin{subfigure}[b]{0.4\textwidth}
    \includegraphics[keepaspectratio=true, height=6.0cm, angle =0]{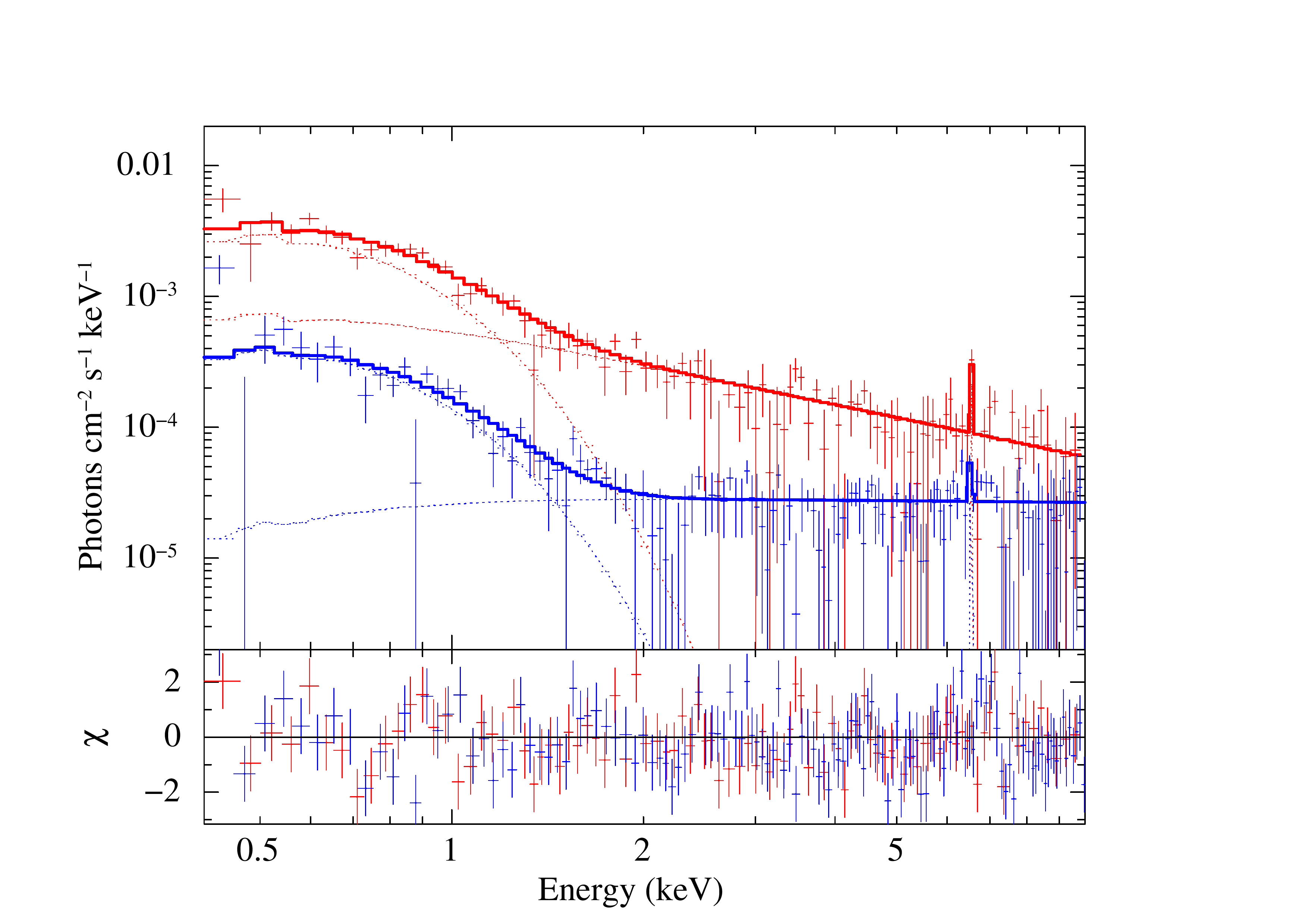}
    \caption{LMC X-4 (XMM-Newton OBSID:203500201)} \label{lmc_x_4_spec}
  \end{subfigure}
  \quad
  \begin{subfigure}[b]{0.4\textwidth}
    \includegraphics[keepaspectratio=true, height=6.0cm, angle =0]{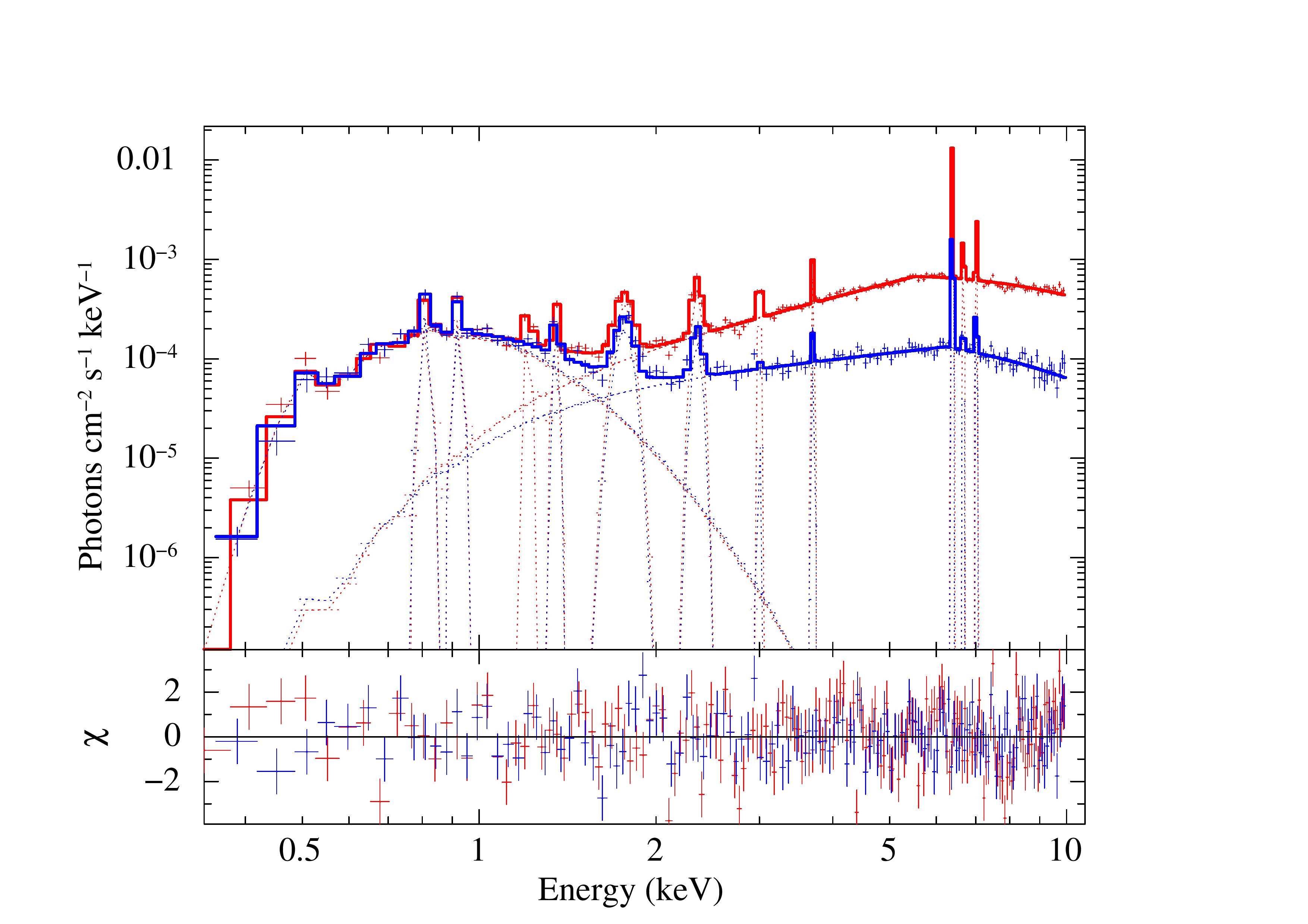}
    \caption{4U 1700-37 (XMM-Newton OBSID:600950101)} \label{4u_1700_37_xmm_spec}
  \end{subfigure}
  \quad
  \begin{subfigure}[b]{0.4\textwidth}
    \includegraphics[keepaspectratio=true, height=6.0cm, angle =0]{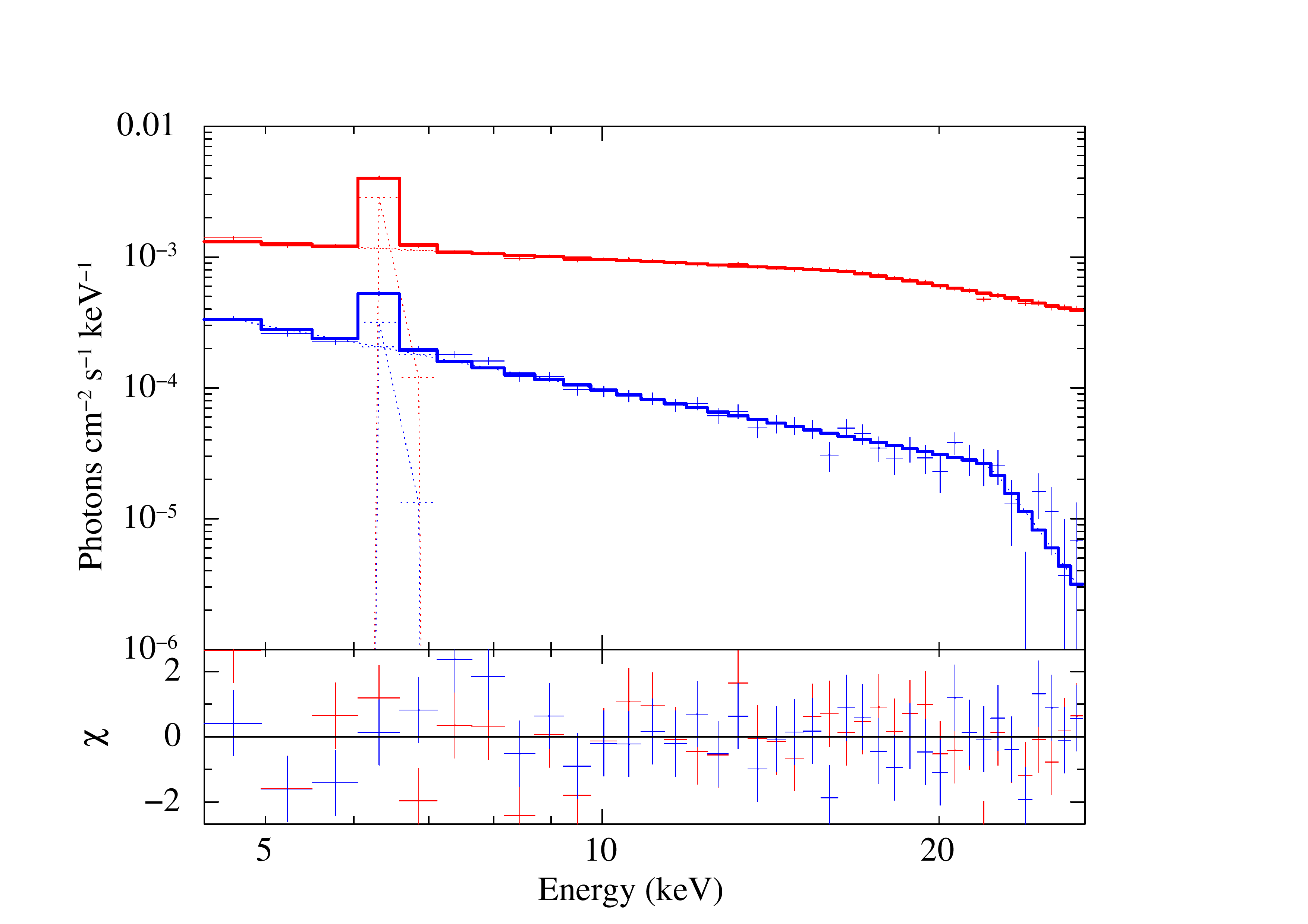}
    \caption{4U 1700-37 (AstroSat OBSID:9000001892)} \label{4u_1700_37_laxpc_spec}
  \end{subfigure}
  \caption{The eclipse flare and eclipse persistent emission duration spectra. The solid red line shows the best fit spectral model for eclipse flare spectrum and the solid blue line shows the same for eclipse persistent emission spectrum. The dotted lines show the individual model components. Both the spectra are fit simultaneously. The bottom panel shows the residuals ($\chi$=(data–model)/error) with respect to the best fit model. The corresponding source name and observation ID are mentioned in the caption below each sub-figure. In the case of \textit{XMM-Newton} spectra of 4U 1700-37 (OBSID: 600950101), the spectral model that models the soft excess with thermal bremsstrahlung (Model 02) is shown.}
  \label{spec_plot}
\end{figure*}


Spectral analysis was performed on the eclipse flare and eclipse persistent emission data by fitting both the spectra simultaneously using \textit{XSPEC Version}: 12.12.1 \citep{xspec1996}. Figure \ref{spec_plot} shows the eclipse flare and eclipse persistent emission spectra of the observations that exhibited flares during the eclipse. The solid red line shows the best fit spectral model for eclipse flare spectrum and the solid blue line shows the same for eclipse persistent emission spectrum. The dotted lines show the individual model components. The bottom panel shows the residuals ($\chi$=(data–model)/error) with respect to the best fit model.

\subsubsection{Vela X-1}
\label{vela_x1_spec_ana}

Multiple emission lines were observed in the 0.5-10 keV \textit{ASCA} SIS eclipse flare and eclipse persistent emission spectra of Vela X-1 (OBSID: 43032000, Figure \ref{vela_x_1_spec}). The spectral continuum was fit with a power-law and a blackbody component was used to fit for the soft excess. Tuebingen-Boulder ISM absorption model (tbabs in XSPEC) was used to account for line of sight absorption. After fitting the spectral continuum, emission features were fit one by one, based on the observed residuals between spectral data and model. While fitting the eclipse flare and eclipse persistent emission spectra simultaneously, the Gaussian line energy parameter of both the spectra were tied to each other. \krtext{The best fit spectral model has the reduced $\chi$\textsuperscript{2} of 1.08 for 450 degrees of freedom. The photon index ($\Gamma$\textsubscript{ph}) of the eclipse flare spectrum is 0.26 and the same for the eclipse persistent emission spectrum is -0.26. The temperature of the blackbody ($kT\textsubscript{bbody}$) is 0.18 keV for the eclipse flare spectrum and 0.20 keV for the eclipse continuum emission spectrum.} Ten emission lines were observed, each of which was fitted with a Gaussian component. We used \textit{AtomDB} (Atomic Data for Astrophysicists) online \textit{WebGUIDE}\footnote[4]{\url{http://www.atomdb.org/index.php}} to identify the detected emission features. The best-fit line energies are observed at 0.91 keV (Ne IX), 1.33 keV (Mg XII), 1.77 keV (Mg XIII), 1.92 keV (Si XIII), 2.37 keV (Si XIV), 2.61 keV (S XVI), 3.07 keV (Ar XVII), 6.68 keV (Fe XXV) and 7.06 keV (Fe XXVI). The line at 6.42 keV is identified as Fe I-XVII line, same as in \cite{nagase1994}. \krtext{The line width of all the narrow emission lines was fixed to 10 eV during spectral fitting.} \cite{nagase1994} have previously analyzed the \textit{ASCA} spectrum of Vela X-1 from different observations (OBSID: 40024000, 40025000), and they too detected many emission lines. The current measurements are also in agreement with those reported in \cite{nagase1994}. \krtext{The unabsorbed X-ray flux is obtained by correcting for the attenuation due to line-of-sight absorption. The unabosrbed model flux in the 2-10 keV energy range for the eclipse flare spectrum is 7.92$\times$10\textsuperscript{-11} ergs cm\textsuperscript{-2} s\textsuperscript{-1} and for the eclipse persistent emission spectrum is 2.65$\times$10\textsuperscript{-11} ergs cm\textsuperscript{-2} s\textsuperscript{-1}. Detailed spectral model and parameters are given in Table \ref{tab:spec_para_velax1_asca}.}

\subsubsection{LMC X-4}
\label{lmc_x4_spec_ana}

The 0.3-10 keV eclipse flare and eclipse persistent emission spectra of LMC X-4 (OBSID: 203500201, Figure \ref{lmc_x_4_spec}) were fitted simultaneously with power-law model to fit for the spectral continuum and a blackbody component to fit for the soft excess. LMC X-4 does not show significant intrinsic absorption. Initially the absorption parameter was allowed to vary however it was not possible to constrain during the spectral fitting. Thus the value of absorption column density was fixed to $\sim$ 8$\times$10\textsuperscript{20} cm\textsuperscript{-2} i.e. the galactic absorption density in the direction of LMC X-4 \citep{hi4pi2016}. \krtext{The reduced $\chi$\textsuperscript{2} of the best fit spectral model is 1.13 for 220 degrees of freedom. The photon index ($\Gamma$\textsubscript{ph}) of the eclipse flare spectrum is 1.02 and the same for the eclipse continuum emission spectrum is 0.05. The blackbody temperatures ($kT_{\text{bbody}}$) for both the eclipse flare spectrum and the eclipse continuum emission spectrum are similar, measuring 0.17 keV and 0.18 keV, respectively. A line observed at 6.55 keV (within the Fe K range of 6.4-7.0 keV) is fit with a Gaussian. While fitting, the Gaussian line energy parameter of both the spectra were tied to each other and the line width was fixed to 10 eV. The unabosrbed model flux in the 2-10 keV energy range for the eclipse flare spectrum is 7.85$\times$10\textsuperscript{-12} ergs cm\textsuperscript{-2} s\textsuperscript{-1} and for the eclipse persistent emission spectrum is 2.09$\times$10\textsuperscript{-12} ergs cm\textsuperscript{-2} s\textsuperscript{-1}. Table \ref{tab:spec_para_lmcx4_xmm} contains the information about the spectral model and the best-fit spectral parameters.}

\subsubsection{4U 1700-37}
\label{4u_1700_37_spec_ana}

Similar to the line dominated spectra of Vela X-1, 0.3-10 keV eclipse flare and eclipse non-flare spectra of 4U 1700-37 (OBSID: 600950101, Figure \ref{4u_1700_37_xmm_spec}) showed multiple emission lines. \krtext{The spectral continuum was fitted by a power-law with high energy cut-off. The photon index ($\Gamma$\textsubscript{ph}) of the eclipse flare spectrum is $\sim$ -1.5 with high energy cut-off at around 5.5 keV and folding energy of around 3.4 keV. The photon index ($\Gamma$\textsubscript{ph}) for the eclipse persistent emission spectrum is around -0.6 with high energy cut-off at $\sim$ 6.2 keV and folding energy of $\sim$ 3.7 keV.} The remarkable feature in this data is a soft excess that did not vary during eclipse flare and eclipse non-flare states. The soft excess can be modelled with a blackbody, bremsstrahlung or with a power-law component \citep{haberl1992,hickox2004,meer2005}. Initially, we modelled the soft excess using the blackbody component. The best-fit blackbody temperature ($kT\textsubscript{bbody}$) was obtained to be 0.20 keV, yielding a reduced $\chi$\textsuperscript{2} value of 1.55 for 281 degrees of freedom. Blackbody temperature component ($kT\textsubscript{bbody}$) in both the spectra were tied together while fitting. The flux of the blackbody component is similar in both spectra indicating that the soft excess does not vary ($Norm\textsubscript{bbody}$ $\sim$ 10\textsuperscript{-5}). This suggests that it is not a reprocessed emission. We also modelled the soft excess using a bremsstrahlung component. \krtext{Fitting with bremsstrahlung yielded a plasma temperature ($kT\textsubscript{bremss}$) of 0.36 keV, with a similar reduced $\chi$\textsuperscript{2} value of 1.54 for 281 degrees of freedom. Plasma temperature component ($kT\textsubscript{bremss}$) in both the spectra were also tied together while fitting. Similar to the blackbody component, the flux of the bremsstrahlung component is also consistent in both the eclipse flare and eclipse persistent emission spectra ($Norm\textsubscript{bremss}$ $\sim$ 5 $\times$10\textsuperscript{-3}).} If the origin of the soft excess is in scattered emission \citep{haberl1992} or in the accretion column \citep{bpaul2002}, the soft excess is expected to be a power-law. The soft excess flux would vary with the overall X-ray flux in such scenario. Since the soft excess flux does not vary during eclipse flare and eclipse non-flare states, we did not model the soft excess with a power-law. Eleven emission lines were identified. Each line was fitted with a Gaussian component in a similar manner as described while fitting the spectra of Vela X-1. The lines were identified using \textit{AtomDB-WebGUIDE}. The best-fit line energies are observed at 0.92 keV (Ne IX), 1.21 keV (Ne X), 1.35 keV (Mg XI), 1.76 keV (Mg XII), 2.35 keV (Si XIII), 3.00 keV (S XV), 3.70 keV (Ca XIX), 6.67 keV (Fe XXV) and 7.02 keV (Fe XXVI). The 0.81 keV line was identified as Ne K\textsubscript{$\alpha$}, and the 6.39 keV line was identified as Fe K\textsubscript{$\alpha$} emission line \citep{thompson2001x}. \krtext{During spectral fitting, the line width for all narrow emission lines was set at 10 eV. The unabosrbed model flux in the 2-10 keV energy range for the eclipse flare spectrum is $\sim$ 51 $\times$10\textsuperscript{-12} ergs cm\textsuperscript{-2} s\textsuperscript{-1} and for the eclipse persistent emission spectrum is $\sim$ 9 $\times$10\textsuperscript{-12} ergs cm\textsuperscript{-2} s\textsuperscript{-1}. Details of both spectral models and their parameters are given in the Table \ref{tab:spec_para_4u170037_xmm}.}

\textit{AstroSat} LAXPC 4-30 keV eclipse flare and eclipse persistent emission spectra of 4U 1700-37 (OBSID: 9000001892, Figure \ref{4u_1700_37_laxpc_spec}) was fitted with an absorbed power-law model with a high energy cut-off. A systematic uncertainty of 1 \% was used during spectral fitting \citep{antia2017,sharma2023}. \krtext{The reduced $\chi$\textsuperscript{2} of the best fit spectral model is \krrtext{1.22} for 76 degrees of freedom.
The absorption column density value,  $\sim$ 5×10\textsuperscript{21} cm\textsuperscript{-2} (i.e., the galactic absorption density in the direction of 4U 1700-37; \citealt{hi4pi2016}), was fixed due to the fitting resulting in a value lower than this. The photon index ($\Gamma$\textsubscript{ph}) of the eclipse flare spectrum is \krrtext{0.43} with high energy cut-off at \krrtext{16.49 keV} and folding energy of \krrtext{20.71 keV}. The photon index ($\Gamma$\textsubscript{ph}) for the eclipse persistent emission spectrum is \krrtext{1.64} with high energy cut-off at \krrtext{22.11 keV} and folding energy of \krrtext{2.45 keV}. Gaussian component was added to model the emission feature. The line energy of the Gaussian is around 6.40 keV which is within the Fe K range of 6.4-7.0 keV. During spectral fitting, the Gaussian line energy parameter was linked between both spectra, while the line width was fixed at 10 eV. The unabosrbed model flux in the 4-30 keV energy range for the eclipse flare spectrum is \krrtext{45.27$\times$10\textsuperscript{-11} ergs cm\textsuperscript{-2} s\textsuperscript{-1}} and for the eclipse persistent emission spectrum is \krrtext{3.04$\times$10\textsuperscript{-11} ergs cm\textsuperscript{-2} s\textsuperscript{-1}}. Table \ref{tab:spec_para_4u170037_astrosat} contains details about the spectral model and parameters.}

In all the spectra, 90\% confidence range was estimated for the spectral parameters using \textit{XSPEC} command \textit{error}. While estimating the errors associated with the emission lines, the line energy and the width were fixed to the best-fit value and only errors associated with normalization parameter were estimated.

\begin{table}
\begin{center}
\caption{Best-fitting parameters for \textit{ASCA} spectrum of Vela X-1. All the errors reported in this table are at 90\% confidence level ($\Delta\chi$\textsuperscript{2}=2.7).}
\label{tab:spec_para_velax1_asca}
\begin{tabular}{ p{1.8cm} p{2.6cm} c c }
\hline
Model	&	Parameters	&	\multicolumn{2}{c}{OBSID: 43032000}\\
\hline
&	& Flare & Non-flare\\
\hline	\\[-1em]	\\[-1em]
TBabs	&	N\textsubscript{H} (10\textsuperscript{22} atoms cm\textsuperscript{-2})	&	$0.76_{-0.14}^{+0.17}$	&	$0.31_{-0.19}^{+0.21}$	\\	\\[-1em]	\\[-1em]
Powerlaw	&	$\Gamma$	&	$0.26_{-0.19}^{+0.19}$	&	$-0.26_{-0.07}^{+0.07}$	\\	\\[-1em]	\\[-1em]
	&	Norm ($\times$10\textsuperscript{-2})	&	$0.14_{-0.03}^{+0.04}$	&	$0.02_{-0.00}^{+0.00}$	\\	\\[-1em]	\\[-1em]
BBody	&	kT\textsubscript{BB} (keV)	&	$0.18_{-0.06}^{+0.06}$	&	$0.20_{-0.04}^{+0.05}$	\\	\\[-1em]	\\[-1em]
	&	Norm ($\times$10\textsuperscript{-3})	&	$0.15_{-0.28}^{+0.28}$	&	$0.01_{-0.01}^{+0.01}$	\\	\\[-1em]	\\[-1em]
Gaussian (1)	&	Energy (keV)	&	0.91	&	0.91	\\	\\[-1em]	\\[-1em]
(Ne IX)	&	Width ($\times$10\textsuperscript{-2} keV)	&	1.00 \textsuperscript{\textit{f}}	&	1.00 \textsuperscript{\textit{f}}	\\	\\[-1em]	\\[-1em]
	&	Norm ($\times$10\textsuperscript{-2})	&	$0.17_{-0.10}^{+0.17}$	&	$0.01_{-0.00}^{+0.00}$	\\	\\[-1em]	\\[-1em]
Gaussian (2)	&	Energy (keV)	&	1.33	&	1.33	\\	\\[-1em]	\\[-1em]
(Mg XII)	&	Width ($\times$10\textsuperscript{-1} keV)	&	0.78	&	0.56	\\	\\[-1em]	\\[-1em]
	&	Norm ($\times$10\textsuperscript{-2})	&	$0.10_{-0.03}^{+0.03}$	&	$0.01_{-0.00}^{+0.00}$	\\	\\[-1em]	\\[-1em]
Gaussian (3)	&	Energy (keV)	&	1.77	&	1.77	\\	\\[-1em]	\\[-1em]
(Mg XII)	&	Width ($\times$10\textsuperscript{-2} keV)	&	1.00 \textsuperscript{\textit{f}}	&	1.00 \textsuperscript{\textit{f}}	\\	\\[-1em]	\\[-1em]
	&	Norm ($\times$10\textsuperscript{-3})	&	$0.31_{-0.10}^{+0.10}$	&	$0.05_{-0.01}^{+0.01}$	\\	\\[-1em]	\\[-1em]
Gaussian (4)	&	Energy (keV)	&	1.92	&	1.92	\\	\\[-1em]	\\[-1em]
(Si XIII)	&	Width ($\times$10\textsuperscript{-2} keV)	&	1.00 \textsuperscript{\textit{f}}	&	1.00 \textsuperscript{\textit{f}}	\\	\\[-1em]	\\[-1em]
	&	Norm ($\times$10\textsuperscript{-3})	&	$0.32_{-0.11}^{+0.11}$	&	$0.02_{-0.01}^{+0.01}$	\\	\\[-1em]	\\[-1em]
Gaussian (5)	&	Energy (keV)	&	2.37	&	2.37	\\	\\[-1em]	\\[-1em]
(Si XIV)	&	Width ($\times$10\textsuperscript{-2} keV)	&	1.00 \textsuperscript{\textit{f}}	&	1.00 \textsuperscript{\textit{f}}	\\	\\[-1em]	\\[-1em]
	&	Norm ($\times$10\textsuperscript{-3})	&	$0.28_{-0.10}^{+0.10}$	&	$0.03_{-0.01}^{+0.01}$	\\	\\[-1em]	\\[-1em]
Gaussian (6)	&	Energy (keV)	&	2.61	&	2.61	\\	\\[-1em]	\\[-1em]
(S XVI)	&	Width ($\times$10\textsuperscript{-2} keV)	&	1.00 \textsuperscript{\textit{f}}	&	1.00 \textsuperscript{\textit{f}}	\\	\\[-1em]	\\[-1em]
	&	Norm ($\times$10\textsuperscript{-4})	&	$0.73_{-0.73}^{+0.75}$	&	$0.06_{-0.06}^{+0.06}$	\\	\\[-1em]	\\[-1em]
Gaussian (7)	&	Energy (keV)	&	3.07	&	3.07	\\	\\[-1em]	\\[-1em]
(Ar XVII)	&	Width ($\times$10\textsuperscript{-2} keV)	&	1.00 \textsuperscript{\textit{f}}	&	1.00 \textsuperscript{\textit{f}}	\\	\\[-1em]	\\[-1em]
	&	Norm ($\times$10\textsuperscript{-3})	&	0.13	&	0	\\	\\[-1em]	\\[-1em]
Gaussian (8)	&	Energy (keV)	&	6.42	&	6.42	\\	\\[-1em]	\\[-1em]
(Fe I-XVII)	&	Width ($\times$10\textsuperscript{-2} keV)	&	1.00 \textsuperscript{\textit{f}}	&	1.00 \textsuperscript{\textit{f}}	\\	\\[-1em]	\\[-1em]
	&	Norm ($\times$10\textsuperscript{-3})	&	$0.88_{-0.19}^{+0.19}$	&	$0.25_{-0.02}^{+0.02}$	\\	\\[-1em]	\\[-1em]
Gaussian (9)	&	Energy (keV)	&	6.68	&	6.68	\\	\\[-1em]	\\[-1em]
(Fe XXV)	&	Width ($\times$10\textsuperscript{-2} keV)	&	1.00 \textsuperscript{\textit{f}}	&	1.00 \textsuperscript{\textit{f}}	\\	\\[-1em]	\\[-1em]
	&	Norm ($\times$10\textsuperscript{-3})	&	$0.23_{-0.17}^{+0.17}$	&	$0.03_{-0.02}^{+0.02}$	\\	\\[-1em]	\\[-1em]
Gaussian (10)	&	Energy (keV)	&	7.06	&	7.06	\\	\\[-1em]	\\[-1em]
(Fe XXVI)	&	Width ($\times$10\textsuperscript{-2} keV)	&	1.00 \textsuperscript{\textit{f}}	&	1.00 \textsuperscript{\textit{f}}	\\	\\[-1em]	\\[-1em]
	&	Norm ($\times$10\textsuperscript{-4})	&	$0.99_{-0.99}^{+1.73}$	&	$0.37_{-0.17}^{+0.17}$	\\	\\[-1em]	\\[-1em]
\hline	\\[-1em]								
\multicolumn{2}{p{4.4cm}}{Reduced $\chi$\textsuperscript{2} / Degrees of freedom}	&	\multicolumn{2}{c}{1.08/450}	\\	\\[-1em]					
\hline	\\[-1em]
\multicolumn{2}{p{4.4cm}}{2-10 keV Unabsorbed model flux (10\textsuperscript{-11} ergs cm\textsuperscript{-2} s\textsuperscript{-1})}&$7.92_{}^{}$&$2.65_{}^{}$\\	\\[-1em]	\\[-1em]
\hline	\\[-1em]	\\[-1em]
\multicolumn{4}{c}{\textsuperscript{\textit{f}}: Parameter was fixed while fitting.}\\
\hline
\hline
\end{tabular}
\end{center}
\end{table}

\begin{table}
\begin{center}
\caption{Best-fitting parameters for \textit{XMM-Newton} spectrum of LMC X-4. All the errors reported in this table are at 90\% confidence level ($\Delta\chi$\textsuperscript{2}=2.7).}
\label{tab:spec_para_lmcx4_xmm}
\begin{tabular}{ p{1.8cm} p{2.6cm} c c }
\hline
Model	&	Parameters	&	\multicolumn{2}{c}{OBSID: 203500201}\\
\hline
&	& Flare & Non-flare\\
\hline	\\[-1em]	\\[-1em]
TBabs	&	N\textsubscript{H} (10\textsuperscript{21} atoms cm\textsuperscript{-2})	&	0.81 \textsuperscript{\textit{f*}}	&	0.81 \textsuperscript{\textit{f*}}	\\	\\[-1em]	\\[-1em]
BBody	&	kT\textsubscript{BB} (keV)	&	$0.17_{-0.01}^{+0.01}$	&	$0.18_{-0.02}^{+0.02}$	\\	\\[-1em]	\\[-1em]
	&	Norm ($\times$10\textsuperscript{-4})	&	$0.38_{-0.04}^{+0.04}$	&	$0.05_{-0.01}^{+0.01}$	\\	\\[-1em]	\\[-1em]
Powerlaw	&	$\Gamma$	&	$1.02_{-0.19}^{+0.18}$	&	$0.05_{-0.18}^{+0.17}$	\\	\\[-1em]	\\[-1em]
	&	Norm ($\times$10\textsuperscript{-3})	&	$0.61_{-0.13}^{+0.15}$	&	$0.03_{-0.01}^{+0.01}$	\\	\\[-1em]	\\[-1em]
Gaussian (1)	&	Energy (keV)	&	6.55	&	6.55	\\	\\[-1em]	\\[-1em]
(Fe K)	&	Width ($\times$10\textsuperscript{-2} keV)	&	1.00 \textsuperscript{\textit{f}}	&	1.00 \textsuperscript{\textit{f}}	\\	\\[-1em]	\\[-1em]
	&	Norm ($\times$10\textsuperscript{-4})	&	$0.21_{-0.14}^{+0.14}$	&	$0.03_{-0.02}^{+0.02}$	\\	\\[-1em]	\\[-1em]
\hline	\\[-1em]
\multicolumn{2}{p{4.2cm}}{Reduced $\chi$\textsuperscript{2} / Degrees of freedom}	&	\multicolumn{2}{c}{1.13/220}	\\	\\[-1em]
\hline	\\[-1em]	\\[-1em]
\multicolumn{2}{p{4.4cm}}{2-10 keV Unabsorbed model flux (10\textsuperscript{-12} ergs cm\textsuperscript{-2} s\textsuperscript{-1})}	&	$7.85_{}^{}$	&	$2.09_{}^{}$\\	\\[-1em]	\\[-1em]
\hline	\\[-1em]	\\[-1em]
\multicolumn{4}{p{8cm}}{\textsuperscript{\textit{f*}}: Parameter was fixed to the value of galactic absorption column density (N\textsubscript{H}) along the line of sight.}\\
\multicolumn{4}{p{8cm}}{\textsuperscript{\textit{f}}: Parameter was fixed while fitting.}\\
\hline
\hline
\end{tabular}
\end{center}
\end{table}

\begin{table*}																
\begin{center}																
\caption{Best-fitting parameters for \textit{XMM-Newton} spectrum of 4U 1700-37. All the errors reported in this table are at 90\% confidence level ($\Delta\chi$\textsuperscript{2}=2.7).}																
\label{tab:spec_para_4u170037_xmm}																
\begin{tabular}{ p{2.8cm} p{3.6cm} c c c c c c }																
\hline																
Model	&	Parameters	&	&	\multicolumn{5}{c}{OBSID: 600950101}\\											
\hline																
	&		&	&	\multicolumn{2}{c}{Model 01 (Bbody)} &   &   \multicolumn{2}{c}{Model 02 (Bremss)}\\						
\hline																
	&		&	&	Flare & Non-flare &   & Flare & Non-flare\\
\hline	\\[-1em]	\\[-1em]
TBabs	&	N\textsubscript{H} (10\textsuperscript{22} atoms cm\textsuperscript{-2})	&	&	$0.47_{-0.09}^{+0.10}$	&	$0.49_{-0.09}^{+0.10}$		&	&	$0.69_{-0.09}^{+0.10}$	&	$0.71_{-0.09}^{+0.09}$	\\	\\[-1em]	\\[-1em]
Bbody/Bremss	&	kT\textsubscript{BB}/kT\textsubscript{bremss} (keV)	&	&	\multicolumn{2}{c}{$0.20_{-0.02}^{+0.02}$}				&	&	\multicolumn{2}{c}{$0.36_{-0.05}^{+0.07}$}	\\	\\[-1em]	\\[-1em]		
	&	Norm ($\times$10\textsuperscript{-4})	&	&	$0.10_{-0.03}^{+0.05}$	&	$0.11_{-0.03}^{+0.05}$		&	&	$47.54_{-21.45}^{+41.47}$	&	$50.54_{-22.47}^{+43.44}$	\\	\\[-1em]	\\[-1em]
HighEcut	&	E\textsubscript{cutoff} (keV)	&	&	$5.48_{-0.10}^{+0.10}$	&	$6.23_{-0.48}^{+0.39}$		&	&	$5.48_{-0.10}^{+0.10}$	&	$6.23_{-0.49}^{+0.39}$	\\	\\[-1em]	\\[-1em]
	&	E\textsubscript{fold} (keV)	&	&	$3.34_{-0.13}^{+0.14}$	&	$3.71_{-0.64}^{+0.76}$		&	&	$3.39_{-0.14}^{+0.15}$	&	$3.72_{-0.64}^{+0.77}$	\\	\\[-1em]	\\[-1em]
Powerlaw	&	$\Gamma$	&	&	$-1.51_{-0.06}^{+0.06}$	&	$-0.63_{-0.11}^{+0.11}$		&	&	$-1.48_{-0.06}^{+0.06}$	&	$-0.62_{-0.12}^{+0.11}$	\\	\\[-1em]	\\[-1em]
	&	Norm ($\times$10\textsuperscript{-4})	&	&	$0.52_{-0.05}^{+0.05}$	&	$0.42_{-0.07}^{+0.07}$		&	&	$0.55_{-0.05}^{+0.05}$	&	$0.43_{-0.07}^{+0.08}$	\\	\\[-1em]	\\[-1em]
Gaussian (1)	&	Energy (keV)	&	&	0.81	&	0.81		&	&	0.81	&	0.81	\\	\\[-1em]	\\[-1em]
(Ne K\textsubscript{$\alpha$})	&	Width ($\times$10\textsuperscript{-2} keV)	&	&	1.00 \textsuperscript{\textit{f}}	&	1.00 \textsuperscript{\textit{f}}		&	&	1.00 \textsuperscript{\textit{f}}	&	1.00 \textsuperscript{\textit{f}}	\\	\\[-1em]	\\[-1em]
	&	Norm ($\times$10\textsuperscript{-4})	&	&	$0.32_{-0.14}^{+0.17}$	&	$0.39_{-0.18}^{+0.23}$		&	&	$0.51_{-0.25}^{+0.30}$	&	$0.64_{-0.32}^{+0.40}$	\\	\\[-1em]	\\[-1em]
Gaussian (2)	&	Energy (keV)	&	&	0.92	&	0.92		&	&	0.92	&	0.92	\\	\\[-1em]	\\[-1em]
(Ne IX)	&	Width ($\times$10\textsuperscript{-2} keV)	&	&	1.00 \textsuperscript{\textit{f}}	&	1.00 \textsuperscript{\textit{f}}		&	&	1.00 \textsuperscript{\textit{f}}	&	1.00 \textsuperscript{\textit{f}}	\\	\\[-1em]	\\[-1em]
	&	Norm ($\times$10\textsuperscript{-4})	&	&	$0.26_{-0.10}^{+0.10}$	&	$0.22_{-0.12}^{+0.13}$		&	&	$0.42_{-0.15}^{+0.17}$	&	$0.35_{-0.19}^{+0.20}$	\\	\\[-1em]	\\[-1em]
Gaussian (3)	&	Energy (keV)	&	&	1.21	&	1.21		&	&	1.21	&	1.21	\\	\\[-1em]	\\[-1em]
(Ne X)	&	Width ($\times$10\textsuperscript{-2} keV)	&	&	1.00 \textsuperscript{\textit{f}}	&	1.00 \textsuperscript{\textit{f}}		&	&	1.00 \textsuperscript{\textit{f}}	&	1.00 \textsuperscript{\textit{f}}	\\	\\[-1em]	\\[-1em]
	&	Norm ($\times$10\textsuperscript{-4})	&	&	0.11	&	0		&	&	0.14	&	0	\\	\\[-1em]	\\[-1em]
Gaussian (4)	&	Energy (keV)	&	&	1.35	&	1.35		&	&	1.35	&	1.35	\\	\\[-1em]	\\[-1em]
(Mg XI)	&	Width ($\times$10\textsuperscript{-2} keV)	&	&	1.00 \textsuperscript{\textit{f}}	&	1.00 \textsuperscript{\textit{f}}		&	&	1.00 \textsuperscript{\textit{f}}	&	1.00 \textsuperscript{\textit{f}}	\\	\\[-1em]	\\[-1em]
	&	Norm ($\times$10\textsuperscript{-4})	&	&	$0.14_{-0.04}^{+0.04}$	&	$0.08_{-0.04}^{+0.05}$		&	&	$0.17_{-0.05}^{+0.05}$	&	$0.10_{-0.05}^{+0.05}$	\\	\\[-1em]	\\[-1em]
Gaussian (5)	&	Energy (keV)	&	&	1.76	&	1.76		&	&	1.76	&	1.76	\\	\\[-1em]	\\[-1em]
(Mg XII)	&	Width ($\times$10\textsuperscript{-1} keV)	&	&	0.52	&	0.52		&	&	0.52	&	0.52	\\	\\[-1em]	\\[-1em]
	&	Norm ($\times$10\textsuperscript{-4})	&	&	$0.56_{-0.05}^{+0.06}$	&	$0.32_{-0.05}^{+0.05}$		&	&	$0.61_{-0.06}^{+0.03}$	&	$0.35_{-0.05}^{+0.05}$	\\	\\[-1em]	\\[-1em]
Gaussian (6)	&	Energy (keV)	&	&	2.35	&	2.35		&	&	2.35	&	2.35	\\	\\[-1em]	\\[-1em]
(Si XIII)	&	Width ($\times$10\textsuperscript{-1} keV)	&	&	0.36	&	0.36		&	&	0.36	&	0.36	\\	\\[-1em]	\\[-1em]
	&	Norm ($\times$10\textsuperscript{-4})	&	&	$0.51_{-0.06}^{+0.06}$	&	$0.16_{-0.04}^{+0.04}$		&	&	$0.53_{-0.06}^{+0.06}$	&	$0.16_{-0.04}^{+0.04}$	\\	\\[-1em]	\\[-1em]
Gaussian (7)	&	Energy (keV)	&	&	3.00	&	3.00		&	&	3.00	&	3.00	\\	\\[-1em]	\\[-1em]
(S XV)	&	Width ($\times$10\textsuperscript{-2} keV)	&	&	1.00 \textsuperscript{\textit{f}}	&	1.00 \textsuperscript{\textit{f}}		&	&	1.00 \textsuperscript{\textit{f}}	&	1.00 \textsuperscript{\textit{f}}	\\	\\[-1em]	\\[-1em]
	&	Norm ($\times$10\textsuperscript{-4})	&	&	$0.22_{-0.05}^{+0.05}$	&	$0.01_{-0.01}^{+0.03}$		&	&	$0.23_{-0.05}^{+0.05}$	&	$0.01_{-0.01}^{+0.03}$	\\	\\[-1em]	\\[-1em]
Gaussian (8)	&	Energy (keV)	&	&	3.70	&	3.70		&	&	3.70	&	3.70	\\	\\[-1em]	\\[-1em]
(Ca XIX)	&	Width ($\times$10\textsuperscript{-2} keV)	&	&	1.00 \textsuperscript{\textit{f}}	&	1.00 \textsuperscript{\textit{f}}		&	&	1.00 \textsuperscript{\textit{f}}	&	1.00 \textsuperscript{\textit{f}}	\\	\\[-1em]	\\[-1em]
	&	Norm ($\times$10\textsuperscript{-4})	&	&	$0.33_{-0.05}^{+0.05}$	&	$0.05_{-0.03}^{+0.03}$		&	&	$0.33_{-0.05}^{+0.05}$	&	$0.05_{-0.03}^{+0.03}$	\\	\\[-1em]	\\[-1em]
Gaussian (9)	&	Energy (keV)	&	&	6.39	&	6.39		&	&	6.39	&	6.39	\\	\\[-1em]	\\[-1em]
(Fe K\textsubscript{$\alpha$})	&	Width ($\times$10\textsuperscript{-2} keV)	&	&	1.00 \textsuperscript{\textit{f}}	&	1.00 \textsuperscript{\textit{f}}		&	&	1.00 \textsuperscript{\textit{f}}	&	1.00 \textsuperscript{\textit{f}}	\\	\\[-1em]	\\[-1em]
	&	Norm ($\times$10\textsuperscript{-3})	&	&	$0.82_{-0.01}^{+0.01}$	&	$0.13_{-0.01}^{+0.01}$		&	&	$0.82_{-0.01}^{+0.01}$	&	$0.13_{-0.01}^{+0.01}$	\\	\\[-1em]	\\[-1em]
Gaussian (10)	&	Energy (keV)	&	&	6.67	&	6.67		&	&	6.67	&	6.67	\\	\\[-1em]	\\[-1em]
(Fe XXV)	&	Width ($\times$10\textsuperscript{-2} keV)	&	&	1.00 \textsuperscript{\textit{f}}	&	1.00 \textsuperscript{\textit{f}}		&	&	1.00 \textsuperscript{\textit{f}}	&	1.00 \textsuperscript{\textit{f}}	\\	\\[-1em]	\\[-1em]
	&	Norm ($\times$10\textsuperscript{-4})	&	&	$0.67_{-0.09}^{+0.09}$	&	$0.06_{-0.05}^{+0.05}$		&	&	$0.67_{-0.09}^{+0.09}$	&	$0.06_{-0.05}^{+0.05}$	\\	\\[-1em]	\\[-1em]
Gaussian (11)	&	Energy (keV)	&	&	7.02	&	7.02		&	&	7.02	&	7.02	\\	\\[-1em]	\\[-1em]
(Fe XXVI)	&	Width ($\times$10\textsuperscript{-2} keV)	&	&	1.00 \textsuperscript{\textit{f}}	&	1.00 \textsuperscript{\textit{f}}		&	&	1.00 \textsuperscript{\textit{f}}	&	1.00 \textsuperscript{\textit{f}}	\\	\\[-1em]	\\[-1em]
	&	Norm ($\times$10\textsuperscript{-3})	&	&	$0.12_{-0.01}^{+0.01}$	&	$0.02_{-0.01}^{+0.00}$		&	&	$0.12_{-0.01}^{+0.01}$	&	$0.02_{-0.01}^{+0.00}$	\\	\\[-1em]	\\[-1em]
\hline	\\[-1em]															
\multicolumn{2}{p{4.4cm}}{Reduced $\chi$\textsuperscript{2} / Degrees of freedom}			&	&		\multicolumn{2}{c}{1.55/281}			&	&		\multicolumn{2}{c}{1.54/281}	\\		\\[-1em]	
\hline	\\[-1em]															
\multicolumn{2}{p{6.4cm}}{2-10 keV Unabsorbed model flux (10\textsuperscript{-12} ergs cm\textsuperscript{-2} s\textsuperscript{-1})}			&	&		$51.41_{}^{}$	&	$9.18_{}^{}$	&	&		$51.63_{}^{}$	&		$9.22_{}^{}$\\	\\[-1em]
\hline	\\[-1em]															
\multicolumn{2}{p{6.4cm}}{\textit{4-10 keV Unabsorbed model flux (10\textsuperscript{-12} ergs cm\textsuperscript{-2} s\textsuperscript{-1})}}			&	&		$48.12_{}^{}$	&	$8.26_{}^{}$	&	&		$48.26_{}^{}$	&		$8.29_{}^{}$\\	\\[-1em]
\hline	\\[-1em]	\\[-1em]
\multicolumn{8}{c}{\textsuperscript{\textit{f}}: Parameter was fixed while fitting.}\\
\hline
\hline
\end{tabular}																
\end{center}																
\end{table*}

\begin{table}
\begin{center}
\caption{Best-fitting parameters for  \textit{Astrosat}-LAXPC spectrum of 4U 1700-37. All the errors reported in this table are at 90\% confidence level ($\Delta\chi$\textsuperscript{2}=2.7).}
\label{tab:spec_para_4u170037_astrosat}
\begin{tabular}{ p{1.8cm} p{2.6cm} c c }
\hline
Model	&	Parameters	&	\multicolumn{2}{c}{OBSID: 9000001892}\\
\hline
&	& Flare & Non-flare\\
\hline	\\[-1em]	\\[-1em]
TBabs	&	N\textsubscript{H} (10\textsuperscript{22} atoms cm\textsuperscript{-2})	&	0.52 \textsuperscript{\textit{f*}}	&	0.52 \textsuperscript{\textit{f*}}	\\	\\[-1em]	\\[-1em]
HighEcut	&	E\textsubscript{cutoff} (keV)	&	$16.49_{-0.70}^{+0.67}$	&	$22.11_{-2.96}^{+3.26}$	\\	\\[-1em]	\\[-1em]
	&	E\textsubscript{fold} (keV)	&	$20.71_{-1.65}^{+1.87}$	&	$2.45_{-2.45}^{+5.48}$	\\	\\[-1em]	\\[-1em]
Powerlaw	&	$\Gamma$	&	$0.43_{-0.04}^{+0.04}$	&	$1.64_{-0.08}^{+0.07}$	\\	\\[-1em]	\\[-1em]
	&	Norm ($\times$10\textsuperscript{-2})	&	$0.26_{-0.02}^{+0.02}$	&	$0.43_{-0.07}^{+0.08}$	\\	\\[-1em]	\\[-1em]
Gaussian (1)	&	Energy (keV)	&	6.40	&	6.40	\\	\\[-1em]	\\[-1em]
(Fe K)	&	Width ($\times$10\textsuperscript{-2} keV)	&	1.00 \textsuperscript{\textit{f}}	&	1.00 \textsuperscript{\textit{f}}	\\	\\[-1em]	\\[-1em]
	&	Norm ($\times$10\textsuperscript{-2})	&	$0.16_{-0.01}^{+0.01}$	&	$0.02_{-0.00}^{+0.00}$	\\	\\[-1em]	\\[-1em]
\hline	\\[-1em]
\multicolumn{2}{p{4.2cm}}{Reduced $\chi$\textsuperscript{2} / Degrees of freedom}	&	\multicolumn{2}{c}{1.22/76}	\\	\\[-1em]
\hline	\\[-1em]	\\[-1em]
\multicolumn{2}{p{4.4cm}}{4-30 keV Unabsorbed model flux (10\textsuperscript{-11} ergs cm\textsuperscript{-2} s\textsuperscript{-1})}	&	$45.27_{}^{}$	&	$3.04_{}^{}$\\	\\[-1em]	\\[-1em]
\hline	\\[-1em]	\\[-1em]
\multicolumn{2}{p{4.4cm}}{\textit{4-10 keV Unabsorbed model flux (10\textsuperscript{-12} ergs cm\textsuperscript{-2} s\textsuperscript{-1})}}	&	$91.63_{}^{}$	&	$14.02_{}^{}$\\	\\[-1em]	\\[-1em]
\hline	\\[-1em]	\\[-1em]
\multicolumn{4}{p{8cm}}{\textsuperscript{\textit{f*}}: Parameter was fixed to the value of galactic absorption column density (N\textsubscript{H}) along the line of sight.}\\
\multicolumn{4}{p{8cm}}{\textsuperscript{\textit{f}}: Parameter was fixed while fitting.}\\
\hline
\hline
\end{tabular}
\end{center}
\end{table}

\section{Discussion}
\label{discussion}

The reprocessed emission observed during the eclipse of HMXB systems provide information about the matter in the environment of the compact object, its material composition and ionization levels, etc. In this section, we discuss about several aspects of the reprocessing region in HMXB systems, based on the results from this analysis.

\subsection{Non-varying soft excess in 4U 1700-37}
\label{bbody_soft_excess}

In addition to continuum power-law emission, an excess in the low energy called soft excess is commonly observed in the X-ray spectra of many HMXBs including 4U 1700-37 \citep{haberl1992, hickox2004, meer2005}. We also observed a soft excess in the eclipse spectrum of 4U 1700-37, carried out with \textit{XMM-Newton} in September 2009 (OBSID 600950101). Remarkably, the soft excess did not vary during eclipse flare and eclipse non-flare states.
We were able to model the soft excess reasonably well using the blackbody and bremsstrahlung components, and obtained similar goodness of fit values for both models. Modelling the soft excess with a blackbody component yielded a blackbody temperature ($kT\textsubscript{bbody}$) of 0.20 keV. The normalization parameter remained almost the same during eclipse flare and eclipse non-flare states. This indicates the origin of the soft excess to be farther from the X-ray source by a distance more than or equal to the radius of the companion star such that its emission is not obstructed during the eclipse and it is not a reprocessed emission. The estimated radius of the blackbody emitting region, obtained from the blackbody normalization parameter, is around 1.2 km. \cite{meer2005} also modeled the soft excess observed in the \textit{XMM-Newton} data of 4U 1700-37, acquired in February 2001, with a blackbody. The radius of the blackbody emitting region estimated by them in the eclipse (0.13 km) and egress states (0.34 km) is small as well. It is difficult to explain the existence of such a small emitting region far away from the X-ray source. The soft excess can also modelled with a power-law component. However, the power-law behavior of the soft excess is often attributed to its origin in either scattered emission \citep{haberl1992} or the accretion column \citep{bpaul2002}. This would result in the change in the soft excess flux with the overall X-ray flux. Since that is not the case, we did not model the soft excess with a power-law. Thermal bremsstrahlung emission can arise in the stellar wind of the companion star. \cite{haberl1992} modelled the soft excess observed in the GINGA observation of 4U 1700-37 using thermal bremsstrahlung ($kT\textsubscript{bremss}$ = 0.46 keV). However, the temperature value of 0.46 keV was obtained during the eclipse ingress state. For the rest of the observations, the temperature could not be constrained and was fixed to 0.46 keV. \krtext{We also modelled the soft excess using thermal bremsstrahlung and obtained a plasma temperature ($kT\textsubscript{bremss}$) of 0.36 keV. The Emission Measure (EM) is defined as $\int n_e n_I dV$, where $n_e$ and $n_I$ represent the electron and ion number densities, respectively.
The EM value of around 5$\times$10\textsuperscript{56} cm\textsuperscript{-3} is obtained from the bremsstrahlung modeling.}

\krtext{Castor, Abbott, Klein (CAK) framework \citep{cak1975} allows estimation of the EM from the stellar and wind paramters. Considering a fully ionized stellar wind composed of hydrogen ions and a radially uniform wind velocity profile $V(r)$ given by ${V_{\infty}}\left(1 - R_{\star}/r\right)$,}

\krtext{
\begin{center}
EM = $\left(\frac{1}{4\pi}\right) \left(\frac{\dot{M}}{m_{H}}\right)^2 \left(\frac{1}{V_{\infty}}\right)^2 \left(\frac{1}{\epsilon R_{\star}} - \frac{1}{r - R_{\star}}\right)$\\
\end{center}
}

\krtext{Where $\dot{M}$ is the mass loss rate from the companion, $m_{H}$ is the mass of the hydrogen atom, $V_{\infty}$ is the terminal wind velocity, $R_{\star}$ is the photospheric radius of the companion star, $\epsilon$ is the distance at which the stellar wind takes off from the photosphere in units of $R_{\star}$, and $r$ is the radius of the emission region. Using stellar and wind parameters calculated by \cite{clark2002} (R\textsubscript{companion} = $21.9_{-0.5}^{+1.3}$ $R_{\odot}$) and \cite{hainich2020} ($\dot{M}\textsubscript{companion}$ = 2.51$\times$10\textsuperscript{-6} $M_{\odot}$ yr\textsuperscript{-1}; $\nu$\textsubscript{$\infty$} = 1.9$\times$10\textsuperscript{3} km s\textsuperscript{-1}), we estimated the size of the region surrounding the companion star capable of generating the observed emission measure.}

We found that the observed bremsstrahlung emission comes from a extremely thin shell around the photosphere of the supergiant companion star, HD 153919. In the framework under consideration, the stellar wind density becomes infinity at the photospheric radius. Thus, considering that the stellar wind takes off just a little distance outside the photosphere ($\epsilon = 0.05$), we find that a thin shell starting from this distance, with a thickness of $\sim$ 10\textsuperscript{-4} times the photospheric radius, can account for the observed bremsstrahlung emission.

\subsection{Line dominated spectrum of Vela X-1 and 4U 1700-37}
\label{emm_spec}

Emission from the X-ray binary systems during the compact object eclipse is a very important probe in understanding the plasma surrounding the compact object via emission line analysis. Due to suppression of continuum spectral component during the eclipse, the emission line components that originate in the surrounding material can be observed with higher contrast. \krtext{Multiple emission lines are observed in the \textit{XMM-Newton} spectrum of 4U 1700-37 (OBSID 600950101) and \textit{ASCA} spectrum of Vela X-1 (43032000). \cite{aftab2019} have previously analyzed the same \textit{XMM-Newton} observation of 4U 1700-37 (OBSID 600950101) and detected fluorescent Fe K\textsubscript{$\alpha$} and highly ionized Fe XXV emission lines along with other low-energy emission lines. Multiple emission lines are also observed in the eclipse spectrum of 4U 1700-37 from HETG (OBSID 18951) on-board \textit{Chandra} \citep{martinezchicharro2021}. They have observed that the K-shell transition lines, corresponding to near neutral atoms, do not diminish much during the eclipse indicating their presence in the extended stellar wind. \cite{martinezchicharro2021} have also found the flux for the highly ionized emission lines (Fe XXV and Fe XXVI Ly$\alpha$) to drop significantly during the eclipse indicating their origin near the X-ray source. \cite{nagase1994} have observed several emission lines in the eclipse spectrum of Vela X-1 using \textit{ASCA} data acquired in 1993 (OBSID: 40024000, 40025000).} We compared the emission line flux of the prominent lines in the eclipse flare spectrum versus eclipse non-flare spectrum. \krtext{In Vela X-1 the Fe I-XVII line flux increases by a factor of $\sim$ 3.5, during the eclipse flare. The Fe K\textsubscript{$\alpha$} line flux increases by about six times during the eclipse flare of \textit{XMM-Newton} observation of 4U 1700-37. Iron line (Fe K) is also observed in the 4-30 keV \textit{AstroSat} spectrum, showing about \krrtext{an eightfold} increase in line flux during the eclipse flare. Unlike the multiple emission lines observed in the \textit{ASCA} spectrum of Vela X-1 and the \textit{XMM-Newton} spectrum of 4U 1700-37, only the Iron line (Fe K) was observed in the \textit{XMM-Newton} spectrum of LMC X-4, showing an increase in flux by a factor of 7 during the eclipse flare.} The increase in the line flux of these lines roughly follows the increase in the overall X-ray flux during the eclipse flare, as expected for reprocessed emission.\\

\subsection{Changing photon index}
\label{cont_spec}

In all three sources showing eclipse flare (Vela X-1, LMC X-4 and 4U 1700-37), the power-law photon index is observed to change during the eclipse flare. Also, the change in the photon index is different for different sources and observations. In \textit{ASCA} observation of Vela X-1 conducted in November 1995 (OBSID 43032000), the photon index ($\Gamma$\textsubscript{ph}) of the eclipse persistent emission spectrum is -0.26 whereas eclipse flare spectrum is softer with $\Gamma$\textsubscript{ph} = 0.26. \cite{martineznunez2014} studied the out-of-eclipse flares in Vela X-1 using \textit{XMM-Newton} data acquired in May 2006 (OBSID 0406430201). They observed no change in the photon index during the out-of-eclipse flare and have reported a constant photon index of 1.60 throughout the observation. In \textit{XMM-Newton} observation of 4U 1700-37 conducted in September 2009 (OBSID 600950101), the eclipse-flare spectrum is harder with $\Gamma$\textsubscript{ph} of $\sim$ -1.5 compared to the $\Gamma$\textsubscript{ph} of $\sim$ -0.6 seen in eclipse persistent emission spectrum. The eclipse flare spectrum of \textit{AstroSat}-LAXPC observation of 4U 1700-37 conducted in February, 2018 (OBSID 9000001892) also shows harder photon index ($\Gamma$\textsubscript{ph} = 0.43) compared to eclipse persistent emission spectrum ($\Gamma$\textsubscript{ph} = 1.64) from the same observation. Unlike eclipse flare spectrum, \cite{boroson2003} found the photon index during out-of-eclipse flare state (0.16) to be higher than the photon index during the out-of-eclipse quiescent state (0.04), in the \textit{Chandra} data of 4U 1700-37 acquired in August 2000. In \textit{XMM-Newton} observation of LMC X-4 conducted in June 2004 (OBSID 203500201), the eclipse flare spectrum has $\Gamma$\textsubscript{ph} of 1.02 which is softer than the eclipse persistent emission spectrum which has a $\Gamma$\textsubscript{ph} of 0.05. The out-of-eclipse persistent emission from the same observation shows intermediate photon index of 0.79.
In the current work, the power-law photon index during the eclipse flares is found to be different, both softer and harder in different sources compared to the non-flare eclipse spectra. It is not obvious that the differences are due to change in the actual emission spectra during flares as observed in out of eclipse durations of these sources.

\subsection{Light-curve variability}
\label{lc_var}

The fastest time variability observed in the X-ray light curve provides an upper limit on the size of the reprocessing region since the variability in the X-ray light curve sharper than the light travel time across the reprocessing region can not observed. \krtext{Flares of a few kilo-second duration were observed in the eclipse light curves of Vela X-1, LMC X-4, and 4U 1700-37, analyzed in this work. Eclipse flares observed by \cite{martinezchicharro2021} also showed similar timescales. In LMC X-4 and 4U 1700-37, the fastest flare rise times, i.e., the minimum count rate doubling times, are 100-200 seconds (See Figure \ref{lc_plot}). This yields a large upper limit on the size of the reprocessing region (100 lightseconds $\approx 43 R_{\odot}$), which is larger than the size of the binary itself (\citealt{falanga2015}, see also Figure \ref{binary_to_scale_vis}).}
Additional constraint on the size of the of reprocessing region comes from the fact that during the X-ray eclipse, the primary radiation is blocked by the companion star and the reprocessed emission only from the structure comparable to, or larger than the size of the companion star can be observed.

\section{Summary and Conclusion}
\label{conclusion}

We have investigated the spectral properties of flares observed during eclipses of three sources: Vela X-1, 4U 1700-37 and LMC X-4. We compared the eclipse flare and eclipse non-flare spectra of these sources. Here are the main findings from this work.

\begin{itemize}
\item In the \textit{XMM-Newton} observation of 4U 1700-37 (OBSID: 600950101), we observed a soft excess in the eclipse spectrum that remained unchanged during both eclipse flare and eclipse non-flare states. Our analysis indicates that this emission is likely originating from an extremely thin shell of the stellar wind surrounding the photosphere of the companion star, HD 153919.
\item During the eclipse of both Vela X-1 (\textit{ASCA} OBSID: 43032000) and 4U 1700-37 (\textit{XMM-Newton} OBSID: 600950101), multiple emission lines were observed in their respective X-ray spectra. During eclipse flares, the increase in the line flux of these lines follows the increase in the overall X-ray flux, as expected for reprocessed emission.
\item We found variations in the power-law photon index of eclipse flare spectra compared to the eclipse non-flare spectra in all three sources. The changes in the photon index differed among the sources and observations, indicating complex spectral behavior during flaring states.
\item \krtext{The fastest flare rise times, i.e., the minimum count rate doubling times, observed in the eclipse flares of LMC X-4 and 4U 1700-37 are around 100 seconds. This gives an upper limit on the size of the reprocessing region of around $43 R_{\odot}$ in these sources.}
\end{itemize}

\section*{Acknowledgements}
\krtext{We thank an anonymous referee for their valuable comments, which have contributed to the improvement of the paper.} This research has made use of archival data and software provided by NASA's High Energy (HEASARC), which is a service of the Astrophysics Science Division at NASA/GSFC. This work has also used the data from the \textit{AstroSat} mission of the Indian Space Research Organisation (ISRO), archived at the Indian Space Science Data Centre (ISSDC). We thank the LAXPC Payload Operation Center (POC) at TIFR, Mumbai for providing necessary software tools. VJ acknowledges the support provided by the Department of Science and Technology (DST) under the “Fund for Improvement of S \& T Infrastructure(FIST)” program (SR/FST/PS-I/2022/208). \krtext{VJ also thanks the Inter-University Centre for Astronomy and Astrophysics (IUCAA), Pune, India, for the Visiting Associateship.}

\section*{Data Availability}
The observational data underlying this work is publicly available through the High Energy Astrophysics Science Archive Research Center (HEASARC). \textit{AstroSat} Data used in this work can be accessed through the Indian Space Science Data Center (ISSDC) at {\url{https://astrobrowse.issdc.gov.in/astro_archive/archive/Home.jsp}}. Any additional information will be shared on reasonable request to the corresponding author.



\bibliographystyle{mnras}
\bibliography{mnras_template} 



\bsp	
\label{lastpage}
\end{document}